\documentclass[a4paper,11pt]{article}
\pdfoutput=1 

\usepackage{jinstpub} 

\usepackage{subcaption}

\usepackage{lineno}

\newcommand{\width}{0.7}

\newcommand{\D}{\mathrm{d}}
\newcommand{\x}{\mathrm{\chi}}
\newcommand{\ex}{\mathrm{x}}
\newcommand{\ey}{\mathrm{y}}
\newcommand{\eps}{\varepsilon}
\newcommand{\prt}{\mathrm{P}} 
\newcommand{\N}{\mathcal{N}}
\newcommand{\tot}{\mathrm{tot}}
\newcommand{\EE}{\mathbb{E}}

\newcommand{\Xmax}{\mathcal{X}}

\newcommand{\Lmax}{\mathcal{L}}
\newcommand{\maxL}{\mathsf{L}_\EE}
\newcommand{\Emx}{\boldsymbol{\mathcal{E}}_{\EE\,}}
\newcommand{\Vmx}{\mathbf{V}_{\EE\,}}
\newcommand{\Wmx}{\boldsymbol{\mathcal{V}}_{\EE\,}}

\title{Study of a data analysis method for the angle resolving silicon telescope}

\author[a,1]{P.~\v{Z}ugec,\note{Corresponding author.}}
\author[b,c]{M.~Barbagallo,}
\author[d]{J.~Andrzejewski,}
\author[d]{J.~Perkowski,}
\author[b]{N.~Colonna,}
\author[a]{D.~Bosnar,}
\author[d]{A.~Gawlik,}
\author[c,e]{M.~Sabat\'{e}-Gilarte,}
\author[c,f]{M.~Bacak,}
\author[c]{F.~Mingrone,}
\author[c]{E.~Chiaveri}
\author[a]{and M.~\v{S}ako}

\affiliation[a]{Department of Physics, Faculty of Science, University of Zagreb,\\Bijeni\v{c}ka cesta 32, 10000 Zagreb, Croatia}
\affiliation[b]{Istituto Nazionale di Fisica Nucleare, Sezione di Bari,\\Via Giovanni Amendola 173, 70126 Bari, Italy} 
\affiliation[c]{European Organization for Nuclear Research (CERN),\\CH-1211 Geneva 23, Switzerland}
\affiliation[d]{Uniwersytet \L\'{o}dzki,\\Ul. Narutowicza 68, 90-136 \L\'{o}d\'{z}, Poland}
\affiliation[e]{Universidad de Sevilla,\\Calle San Fernando, 4, 41004 Sevilla, Spain}
\affiliation[f]{Technische Universit\"{a}t Wien,\\Karlsplatz 13, 1040 Wien, Austria}

\emailAdd{pzugec@phy.hr}

\collaboration[c]{on behalf of n\_TOF collaboration}

\abstract{A new data analysis method is developed for the angle resolving silicon telescope introduced at the neutron time of flight facility n\_TOF at CERN. The telescope has already been used in measurements of several neutron induced reactions with charged particles in the exit channel. The development of a highly detailed method is necessitated by the latest joint measurement of the $^{12}$C($n,p$) and $^{12}$C($n,d$) reactions from n\_TOF. The reliable analysis of these data must account for the challenging nature of the involved reactions, as they are affected by the multiple excited states in the daughter nuclei and characterized by the anisotropic angular distributions of the reaction products. The unabridged analysis procedure aims at the separate reconstruction of all relevant reaction parameters --- the absolute cross section, the branching ratios and the angular distributions --- from the integral number of the coincidental counts detected by the separate pairs of silicon strips. This procedure is tested under the specific conditions relevant for the $^{12}$C($n,p$) and $^{12}$C($n,d$) measurements from n\_TOF, in order to assess its direct applicability to these experimental data. Based on the reached conclusions, the original method is adapted to a particular level of uncertainties in the input data.}


\begin{document}
\maketitle
\flushbottom

\section{Introduction}
\label{introduction}

The neutron time of flight facility n\_TOF at CERN is a neutron production facility aiming at measuring the neutron induced reactions. A massive lead spallation target irradiated by the 20~GeV proton beam from the CERN Proton Synchrotron serves as the primary source of neutrons, delivering an extremely luminous white neutron beam spanning 12 orders of magnitude in energy --- from 10~meV to 10~GeV. The n\_TOF facility features two experimental areas: Experimental Area~1 (EAR1), horizontally placed at 185~m from the spallation target, and the Experimental Area~2 (EAR2) vertically placed at 20~m above the target. While EAR1 is best adjusted to the high neutron energy and the high resolution measurements, EAR2 excels at the measurements with small, highly radioactive samples characterized by low cross sections for the investigated reactions. More details on the general features of the n\_TOF facility and EAR1 itself may be found in ref.~\cite{ntof}, while the specifics on EAR2 are addressed in Refs.~\cite{ear2_1,ear2_2,ear2_3}. An overview of the experimental program at n\_TOF may be found in ref.~\cite{ntof_rev}. A general overview of many different types of detectors used at n\_TOF for the measurements of various types of the neutron induced reactions, together with the detailed description of the procedures for the analysis of electronic signals from these detectors, can be found in ref.~\cite{psa}.

\begin{figure}[t!]
\begin{subfigure}{0.45\textwidth}
\centering
  \includegraphics[height=0.3\textheight]{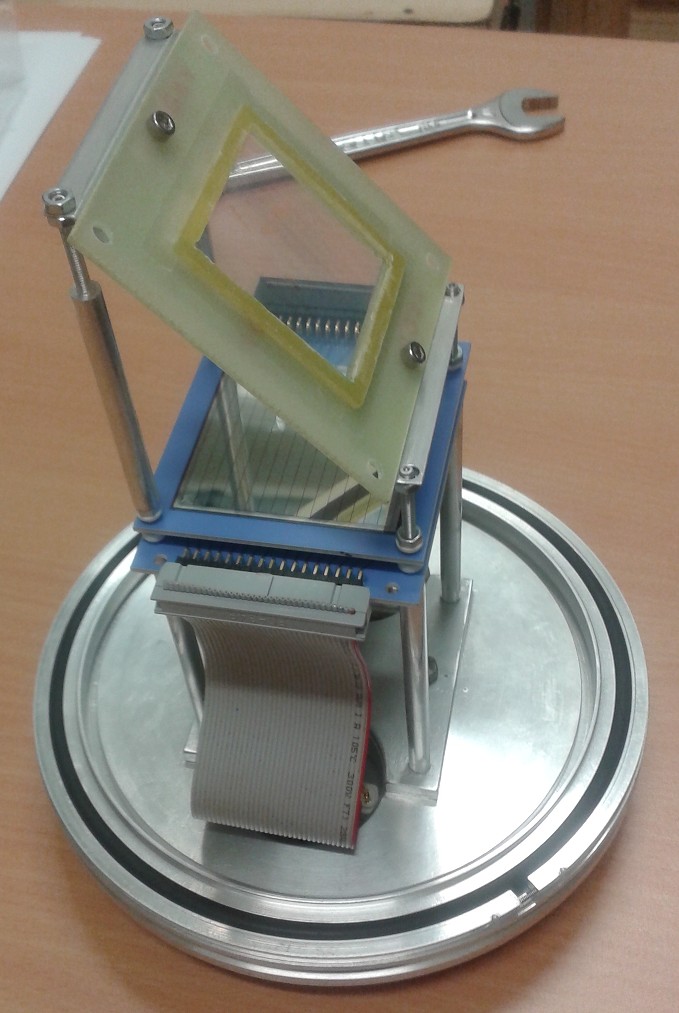}  
\caption{}
\label{fig0_a}
\end{subfigure}
\begin{subfigure}{.5\textwidth}
\centering
  \includegraphics[height=0.3\textheight]{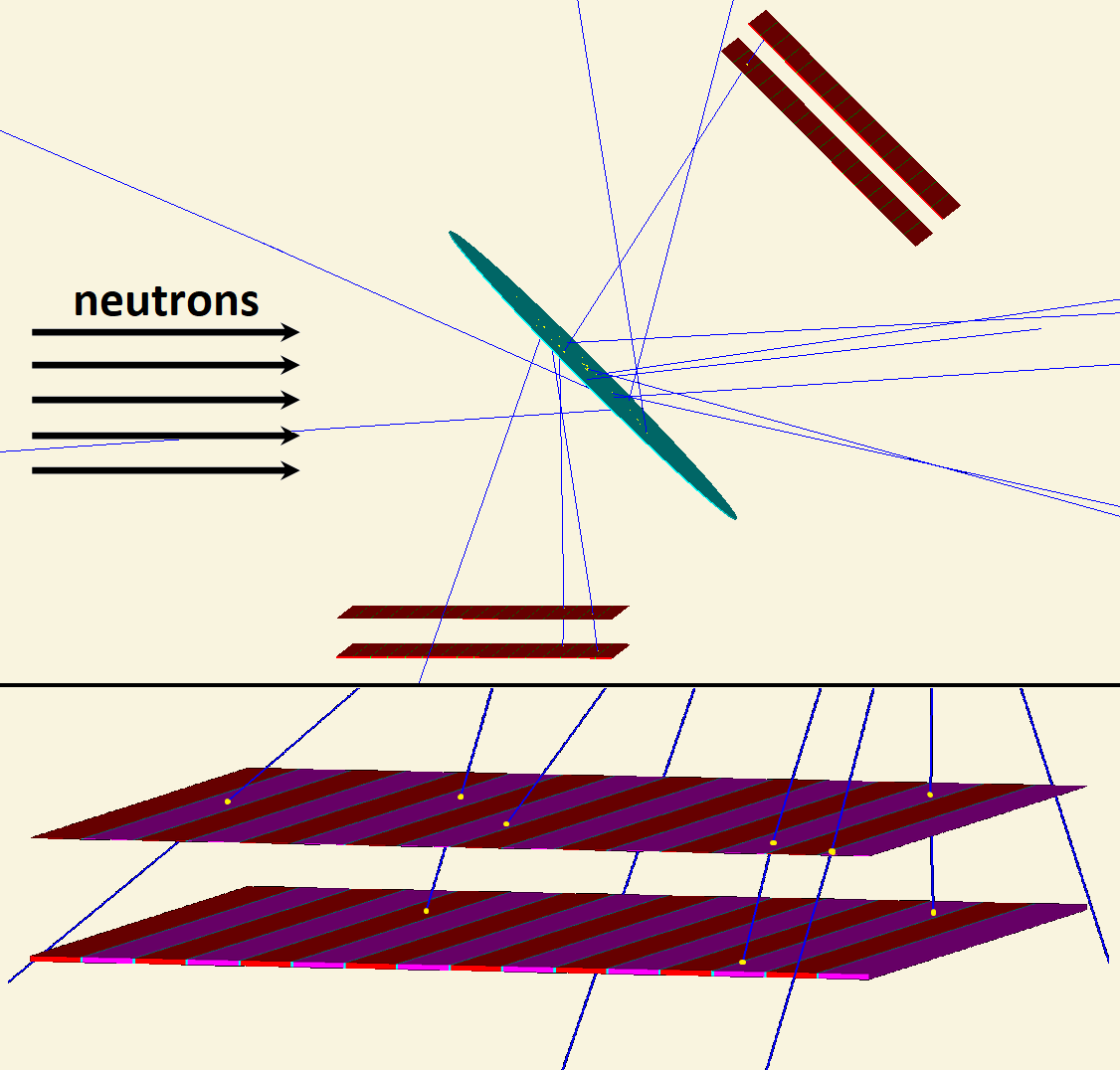}  
\caption{}
\label{fig0_b}
\end{subfigure}
\caption{(a) Experimental setup housing the segmented silicon telescope (SITE), originally used for the measurement of the $^7$Be($n,p$) reaction. (b) Top: upgraded experimental configuration used for the measurements of the $^{12}$C($n,p$) and $^{12}$C($n,d$) reactions, comprising two silicon telescopes (a caption from Geant4 simulations). The central object is a carbon sample, as a source of several displayed proton trajectories. Bottom: closeup of a rear telescope, showing a stripped structure of $\Delta \mathrm{E}$ and $\mathrm{E}$ layers (the strips of alternating colors, separated by a very thin layer of inactive silicon).}
\label{fig0}
\end{figure}


A new, highly sophisticated silicon telescope (SITE) has recently been introduced at n\_TOF for measurements of the neutron induced reactions with charged particles in the exit channel \cite{site_np}. It consists of two separate and segmented layers of 16 silicon strips, 5~cm~$\times$~3~mm each, placed in parallel between the layers. The detector is shown in figure~\ref{fig0_a}. Both layers are 5~cm~$\times$~5~cm in lateral dimensions, distanced by 7~mm. The first ($\Delta \mathrm{E}$) layer and the second ($\mathrm{E}$) layer are 20~$\mu$m and 300~$\mu$m thick, respectively. The detector allows to discriminate different types of charged particles using the $\Delta \mathrm{E}$-$\mathrm{E}$ telescope method, while also offering the limited angular discrimination, governed by its geometry and the sample-relative positioning.

Excellent particle discrimination capabilities of this silicon telescope have been clearly demonstrated \cite{site_np} and it has already been successfully used in the challenging measurement of the $^7$Be($n,p$) reaction cross section, highly relevant for the long-standing Cosmological Lithium Problem \cite{be_np}. This measurement has also been accompanied by the measurement of the $^7$Be($n,\alpha$) reaction cross section \cite{be_na}, employing a similar type of silicon sandwich detector \cite{sandw_na}.


Rather recently an integral measurement of the $^{12}$C($n,p$)$^{12}$B reaction has been performed at n\_TOF, using two liquid scintillation C$_6$D$_6$ detectors for the detection of $\beta$-rays from the decay of the produced $^{12}$B \cite{carbon_prc,carbon_epja}. The results of this measurement have turned out somewhat surprising, lying entirely outside of values predicted by all earlier datasets available for this reaction (experimental or otherwise), which are in a rather poor agreement between each other (for a concise review of these datasets see Refs.~\cite{carbon_prc,carbon_epja,carbon_prop}). In order to resolve this conundrum a more advanced energy-differential measurement of the $^{12}$C($n,p$)$^{12}$B and $^{12}$C($n,d$)$^{11}$B reactions was proposed \cite{carbon_prop} and already performed at EAR1 at n\_TOF, using an upgraded SITE configuration displayed in figure~\ref{fig0_b}. The upgrade consisted in introducing a second telescope in order to increase the angular coverage as much as possible, while keeping both of them outside of the neutron beam. We will refer to these two telescopes as \textit{front} and \textit{rear}, the front one being parallel to the sample and covering the forward angles, with the rear one being parallel to the neutron beam and mostly covering the backward angles (see figure~\ref{fig0_b}). The analysis of the experimental data on the $^{12}$C($n,p$) and the $^{12}$C($n,d$) reaction is under way, pending the development of a new analysis method for extracting the relevant reaction parameters. Most important among these is the absolute cross section. The angle-differential cross sections for the reaction flow via the separate excited states of a daughter nucleus ($^{11}$B \cite{{b11_states}} or $^{12}$B \cite{b12_states}) would also be highly desirable. However, the reliable decoupling of these states might not be possible at the level of statistics expected from the latest measurement.

The purpose of this paper is twofold. The first is to develop the necessary formalism for the analysis of the data obtained with the multi-channel telescope (section~\ref{derivation}). The second is to investigate its applicability on the artificially generated dataset resembling the first experimental dataset from n\_TOF to which the the method is to be applied at a later date: that of the $^{12}$C($n,p$) reaction (section~\ref{demonstration}). In doing this we aim (1) to provide the future users of the method with all the necessary steps and considerations to be taken into account in extracting the optimal set of physical parameters from a given measurement; (2) to provide an honest assessment of the \emph{direct} applicability of the method to a dataset of a given level of uncertainties, in particular the one expected from the $^{12}$C($n,p$) measurement, and (3) to provide alternative solutions in case the \text{direct} application proves to be unreliable due to the level of uncertainties in the extracted results (section~\ref{reduction}). Since we develop the method having a specific $^{12}$C($n,p$) reaction in mind, it cannot be overemphasized that the procedure is aimed at and designed for the particular detector setup, based on the $\Delta \mathrm{E}$-$\mathrm{E}$ telescoping principle, rather than for the particular reaction of even the type of reaction. Therefore, at no point should the method be considered as limited to this specific (type of) reaction, nor should the conclusions regarding a particular $^{12}$C($n,p$) measurement from n\_TOF be misinterpreted for some general limitation of the method itself.


\section{Method derivation}
\label{derivation}


Let $\theta$ be the angle of proton emission in the \textit{center of mass} frame (of the incoming neutron and $^{12}$C nucleus before the reaction, and of the outgoing proton and the $^{12}$B nucleus after the reaction), relative to the direction of the neutron beam. We immediately introduce:
\begin{linenomath}\begin{equation}
\x\equiv\cos\theta
\end{equation}\end{linenomath}
as a relevant variable. For simplicity of terminology we still refer to $\x$ as the \textit{angle} of the proton emission. Let $N_{ij}$ be the \textit{total} number of protons detected in coincidence by the \mbox{$(i,j)$-pair} of strips, with the first index $i$ denoting some of the thin $\Delta \mathrm{E}$-strips and the a second index $j$ denoting some of the thick $\mathrm{E}$-strips from any telescope (front or rear). Let $E$ be the energy of the incident neutron. The proton produced by the neutron of sufficiently high energy might be emitted leaving the $^{12}$B nucleus in any of the energetically accessible states. Thus, the proton energy is clearly contingent on the daughter nucleus' excited states. Denoting these states by $\ex$ ($\ex=0$ being the ground state), we define the probability $\eps_{ij}(\ex,E,\x)$ for the coincidental detection --- by the $(i,j)$-pair of strips --- of protons produced by the neutrons of energy $E$ and emitted under an angle $\x$ leaving the $^{12}$B nucleus in a state $\ex$:
\begin{linenomath}\begin{equation}
\eps_{ij}(\ex,E,\x)\equiv\frac{\D^2 N_{ij}(\ex,E,\x)}{\D^2 \N(\ex,E,\x)},
\label{epsilon}
\end{equation}\end{linenomath}
with $\D^2 N_{ij}(\ex,E,\x)$ as the number of the detected protons and $\D^2 \N(\ex,E,\x)$ as the number of protons emitted under such conditions. These probabilities may easily be obtained from the dedicated simulations, described in appendix~\ref{app_eff}. It should be noted that they reflect the properties of the experimental setup itself, and are independent of the angular distribution of the emitted protons.

\begin{figure}[t!]
\centering
\includegraphics[width=\width\linewidth]{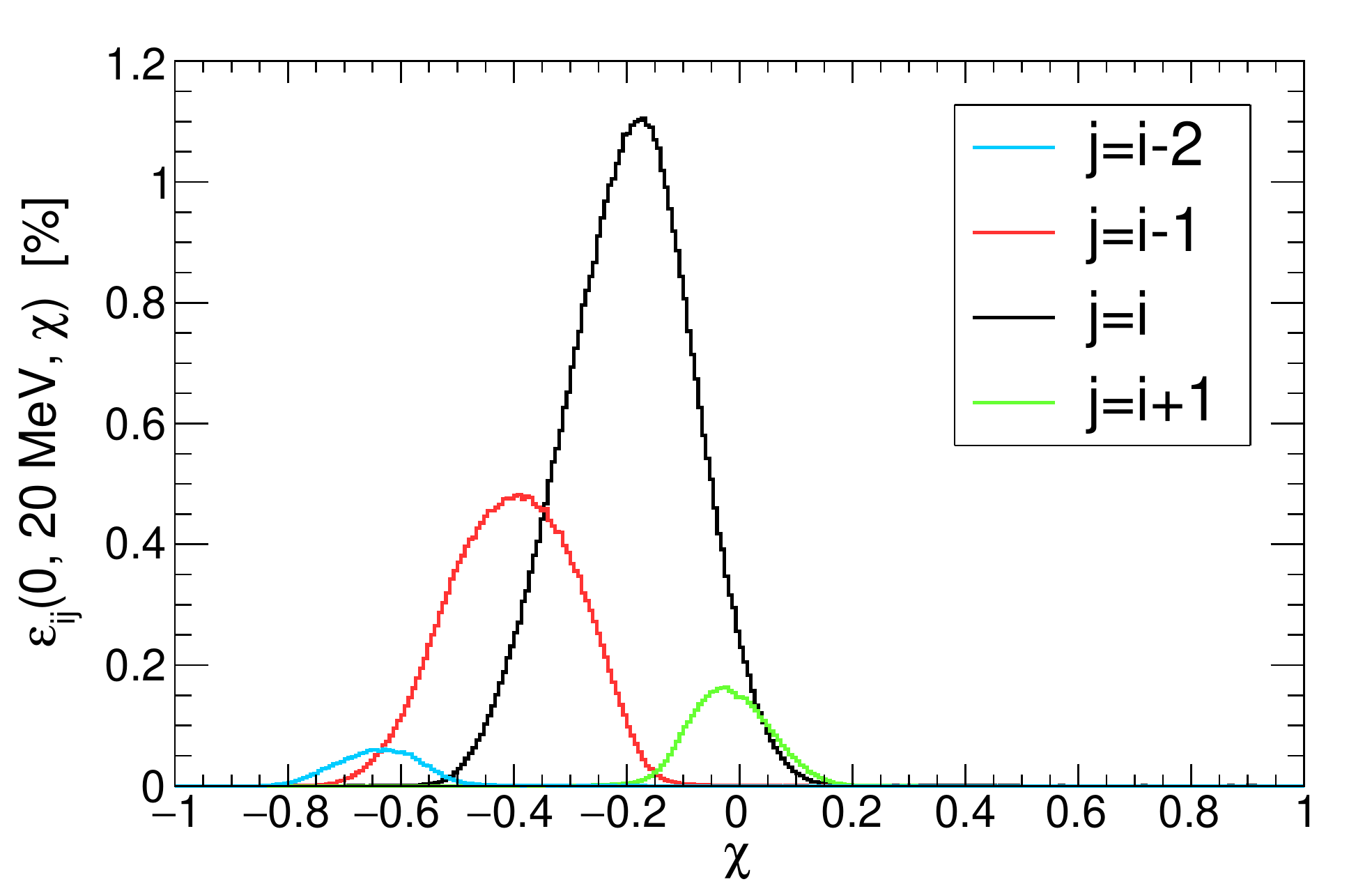}
\caption{Examples of the coincidental detection probabilities for protons from the $^{12}$C($n,p$) reaction induced by $20$~MeV neutrons, leaving the $^{12}$B nucleus in the ground state. The probabilities are shown for an arbitrary ($i$-th) $\Delta \mathrm{E}$-strip in coincidence with the several closest $\mathrm{E}$-strips.}
\label{fig3}
\end{figure}

Only for illustration purposes, figure~\ref{fig3} shows the coincidental detection probabilities \linebreak $\eps_{ij}(0,20\:\mathrm{MeV},\x)$ for protons produced by 20~MeV neutrons, leaving the $^{12}$B nucleus in the ground state (\mbox{$\ex=0$}). The probabilities are shown for some arbitrarily selected, $i$-th $\Delta \mathrm{E}$-strip in coincidence with the several closest $\mathrm{E}$-strips. During the method implementation these curves, i.e. their smooth(ed) forms never have to be constructed, as their integrals can be calculated as the weighted sum of the simulated counts. The issue is further addressed in appendix~\ref{app_eff}.


The number of protons emitted under the described specific conditions may be decomposed as:
\begin{linenomath}\begin{equation}
\D^2 \N(\ex,E,\x)=\phi(E)\mu(E)\frac{\varrho(\ex,E,\x)}{\Sigma_\tot(E)}\left(1-e^{-\eta\Sigma_\tot(E)}\right)\D E \D \x,
\label{N_prod}
\end{equation}\end{linenomath}
with $\phi(E)$ as a time-integrated energy dependent neutron flux irradiating the sample: \linebreak \mbox{$\phi(E)=\D \Phi(E)/\D E$}, $\D \Phi(E)$ being the total number of neutrons of energy $E$ intercepted by the sample. The multiple scattering factor $\mu(E)$ describes an increase in the  neutron flux at an energy $E$ due to the energy loss of higher-energy neutrons by means of the multiple scattering inside the sample itself. With $\varrho(\ex,E,\x)$ as the partial cross section for the $^{12}$C($n,p$) reaction, i.e. for a particular reaction of interest, $\Sigma_\tot(E)$ is the total cross section for \textit{any} neutron induced reaction in the carbon sample. Finally, $\eta$ is the areal density of the sample, as encountered by the neutron beam, in the number of atoms per unit area. While the term \mbox{$1-e^{-\eta\Sigma_\tot(E)}$} gives a probability for any neutron reaction to occur (the exponential term itself being the transmission probability), the differential ratio \mbox{$\varrho(\ex,E,\x)/\Sigma_\tot(E)$} governs the probability of that reaction being the one of interest. The differential cross section may now be decomposed as:
\begin{linenomath}\begin{equation}
\varrho(\ex,E,\x)=\sigma(E)\,\rho(\ex,E)\,A(\ex,E,\x),
\label{decomp}
\end{equation}\end{linenomath}
with $\sigma(E)$ as the total cross section for the $^{12}$C($n,p$) reaction, $\rho(\ex,E)$ as the energy-dependent branching ratios for the reaction flow via the particular excited state of $^{12}$B, and $A(\ex,E,\x)$ as the angular distribution of protons specific to that state.


From eq.~(\ref{N_prod}) we now isolate all the terms that are independent of the detector setup, while being available from the experiment, simulation or any evaluation database:
\begin{linenomath}\begin{equation}
w(E)\equiv\frac{1-e^{-\eta\Sigma_\tot(E)}}{\Sigma_\tot(E)}\phi(E)\mu(E).
\label{w_exact}
\end{equation}\end{linenomath}
The neutron flux $\phi(E)$ at EAR1 (as well as the flux at EAR2 \cite{flux_ear2}) is available from the dedicated measurements at n\_TOF \cite{flux_ear1}. Even in a general case of a thick sample, the multiple scattering factor could be obtained from the dedicated simulations if the total cross section $\Sigma_\tot(E)$ and the elastic scattering cross section $\Sigma_\mathrm{el}(E)$ for carbon were known with sufficient precision, which they are \cite{carbon_epja}. However, as the very thin carbon sample was used during the energy-differential measurement --- 0.25~mm \cite{carbon_prop}, with the thickness of 0.35~mm, i.e. an areal density of $\eta=4\times10^{-3}$~atoms/barn being intercepted by the neutron beam due to the sample tilt of $45^\circ$ (figure~\ref{fig0_b}) --- a thin sample approximation becomes highly appropriate. In this approximation the deviation of the multiple scattering factor from unity is completely negligible: \mbox{$\mu(E)\approx1$}, while the full fractional term from eq.~(\ref{w_exact}) approximates to $\eta$ due to \mbox{$\eta\Sigma_\tot(E)\ll1$} within the entire neutron energy range of interest. Hence:
\begin{linenomath}\begin{equation}
w(E)\approx\eta\phi(E).
\label{w_approx}
\end{equation}\end{linenomath}
Using Eqs.~(\ref{decomp}) and (\ref{w_exact}), eq.~(\ref{N_prod}) may now be rewritten as:
\begin{linenomath}\begin{equation}
\D^2 \N(\ex,E,\x)=w(E)\,\sigma(E)\,\rho(\ex,E)\,A(\ex,E,\x)\,\D E \D \x.
\label{decomposed}
\end{equation}\end{linenomath}
We now take into account that due to the energy spread of the neutron beam the experimental data must be analyzed within the energy intervals of finite width. We use the following notation for one such interval:
\begin{linenomath}\begin{equation}
\EE\equiv [E_\mathrm{min},E_\mathrm{max}],
\label{dE}
\end{equation}\end{linenomath}
meaning that all the later quantities denoted by $\EE$ are either integrals or averages over $\EE$, or that they may be separately and independently selected for each such energy interval. Since any particular method implementation requires a weighted averaging over $w(E)$, we immediately introduce the following norm:
\begin{linenomath}\begin{equation}
W_\EE\equiv\int_\EE w(E)\D E.
\end{equation}\end{linenomath}
Returning to the differential number of protons $\D^2 N_{ij}(\ex,E,\x)$ detected by a particular pair of strips and recalling that there may be multiple excited states of the daughter nucleus contributing to the reaction, we may write the expression for the total number of protons detected by the ($i,j$)-pair of strips:
\begin{linenomath}\begin{equation}
N_{ij}^{(\EE)}=\sum_{\ex=0}^{\Xmax_\EE}\int_\EE \D E \int_{-1}^1 \D\x \: \frac{\D^2 N_{ij}(\ex,E,\x)}{\D E \D \x},
\label{master_0}
\end{equation}\end{linenomath}
with $\Xmax_\EE$ as the highest excited state affecting the data from the energy interval $\EE$. It should be noted that the total detected counts $N_{ij}^{(\EE)}$ taken for analysis will also be dependent on the \textit{energy deposition cuts} imposed on the experimental data. We will consider this dependence \textit{implicitly absorbed} within the terms $N_{ij}^{(\EE)}$ and, consequently, the corresponding detection probabilities $\eps_{ij}(\ex,E,\x)$.

We now define an arbitrary bijective mapping:
\begin{linenomath}\begin{equation}
(i,j)\mapsto\alpha,
\label{alpha}
\end{equation}\end{linenomath}
allowing us to write eq.~(\ref{master_0}) in a single index, which will soon become useful in bringing the system to an appropriate matrix form. This bijection never needs to be explicitly constructed. Using this formal manipulation in conjunction with applying Eqs.~(\ref{epsilon}) and (\ref{decomposed}) to eq.~(\ref{master_0}), we arrive at the master equation:
\begin{linenomath}\begin{equation}
N_\alpha^{(\EE)}=\sum_{\ex=0}^{\Xmax_\EE}\int_\EE \D E \int_{-1}^1 \D\x \; \eps_\alpha(\ex,E,\x)w(E)\sigma(E)\rho(\ex,E)A(\ex,E,\x).
\label{master}
\end{equation}\end{linenomath}
Our goal is now to bring this equation into the matrix form:
\begin{linenomath}\begin{equation}
\vec{N}^{(\EE)}=\Emx \vec{\prt}^{\,(\EE)}
\label{vectorized}
\end{equation}\end{linenomath}
by constructing the vector $\vec{N}^{(\EE)}$ of total detected counts $N_\alpha^{(\EE)}$, by designing an appropriate matrix $\Emx$ and by isolating the sought parameters of the partial cross section within the vector $\vec{\prt}^{\,(\EE)}$. We will obtain this matrix form by decomposing the angular distributions into partial waves (Legendre polynomials).

Before proceeding further let us put forth the tools and considerations common to any particular implementation of the method. Let us denote by $\mathcal{R}_\EE$ the number of \textit{relevant pairs of strips} composing the experimental dataset from $\vec{N}^{(\EE)}$ and by $\mathcal{P}_\EE$ the number of the partial cross section parameters from $\vec{\prt}^{\,(\EE)}$. Then we can select at most $\mathcal{R}_\EE$ parameters to reconstruct:
\begin{linenomath}\begin{equation}
\left.\begin{array}{l}
\mathcal{R}_\EE\equiv\dim\big[\vec{N}^{(\EE)}\big]\\
\mathcal{P}_\EE\equiv\dim\big[\vec{\prt}^{\,(\EE)}\big]
\end{array}\right\}
\quad\Rightarrow\quad
\mathcal{P}_\EE\le\mathcal{R}_\EE.
\label{constraint}
\end{equation}\end{linenomath}
When $\mathcal{P}_\EE<\mathcal{R}_\EE$, the best solution to this system may be found by means of the weighted least squares method \cite{data_book}:
\begin{linenomath}\begin{equation}
\vec{\prt}^{\,(\EE)} =\big(\Emx^\top\Vmx^{-1}\Emx\big)^{-1} \Emx^\top\Vmx^{-1} \vec{N}^{(\EE)},
\label{lsq_mean}
\end{equation}\end{linenomath}
with $\Vmx$ are the covariance matrix of the input data, allowing for the propagation of experimental uncertainties in order to obtain the covariance matrix $\Wmx$ of the reconstructed parameters and their respective uncertainties $\delta\prt^{(\EE)}_\beta$ as:
\begin{linenomath}\begin{equation}
\Wmx=\big(\Emx^\top\Vmx^{-1}\Emx\big)^{-1} \quad\Rightarrow\quad \delta\prt^{(\EE)}_\beta=\sqrt{[\Wmx]_{\beta\beta}}.
\label{lsq_uncer}
\end{equation}\end{linenomath}
From the raw results obtained from eq.~(\ref{lsq_mean}) we will have to calculate various consequent quantities --- the total reaction cross section, branching ratios and the angular distribution parameters --- while dealing with the high uncertainties and possible correlations in the reconstructed $\vec{\prt}^{\,(\EE)}$. Therefore, we are well advised to take into account the effects of the full covariance matrix upon the propagation of uncertainties. Let any of these quantities be the scalar function of $\vec{\prt}^{\,(\EE)}$ that we generally denote as: \mbox{$F_\EE\equiv F\big(\vec{\prt}^{\,(\EE)}\big)$}. Then the respective uncertainty $\delta F_\EE$ may be expressed as: 
\begin{linenomath}\begin{equation}
\delta F_\EE=\sqrt{\mathbf{J}_F \Wmx \mathbf{J}_F^\top}=\sqrt{\sum_{\beta=1}^{\mathcal{P}_\EE}\sum_{\beta'=1}^{\mathcal{P}_\EE} \frac{\partial F_\EE}{\partial \prt_{\beta}^{(\EE)}} \frac{\partial F_\EE}{\partial \prt_{\beta'}^{(\EE)}} [\Wmx]_{\beta\beta'}},
\label{uncer}
\end{equation}\end{linenomath}
with $\mathbf{J}_F$ indicating the conventionally defined Jacobian matrix of the function $F$.



\subsection{Partial waves decomposition}
\label{pw_decomposition}

We start by decomposing the angular distributions into the selected number of partial waves, i.e. Legendre polynomials $P_l(\x)$:
\begin{linenomath}\begin{equation}
A(\ex,E,\x)\approx\sum_{l=0}^{\Lmax_\EE(\ex)} a_l(\ex,E)\, P_l(\x),
\label{A_decomp}
\end{equation}\end{linenomath}
with the maximum wave number $\Lmax_\EE(\ex)$ freely adjustable to any given excited state, within the constraints of the total number $\mathcal{R}_\EE$ of the available data points. Eq.~(\ref{master}) may now be rewritten as:
\begin{linenomath}\begin{align}
\begin{split}
N_\alpha^{(\EE)}&\approx\sum_{\ex=0}^{\Xmax_\EE}\sum_{l=0}^{\Lmax_\EE(\ex)}  \int_\EE  w(E)\,\sigma(E)\,\rho(\ex,E)\,a_l(\ex,E)\,\D E\,\int_{-1}^1\eps_\alpha(\ex,E,\x) P_l(\x)\,\D\x \\
&=2W_\EE \sum_{\ex=0}^{\Xmax_\EE}\sum_{l=0}^{\Lmax_\EE(\ex)}  \big\langle \sigma\rho a_l\, \big\langle \eps_\alpha \big\rangle_{l\,} \big\rangle_{\EE},
\end{split}
\label{recast_Pl}
\end{align}\end{linenomath}
where we recognize the appearance of the weighted averages $\langle\cdot\rangle_\EE$ and $\langle\cdot\rangle_l$, with $w(E)$ and $P_l(\x)$ as the respective weighting functions. We now approximate the average product by the product of averages:
\begin{linenomath}\begin{equation}
\big\langle \sigma\rho a_l\, \big\langle \eps_\alpha \big\rangle_{l\,} \big\rangle_{\EE} \approx \bar{\bar{\eps}}_{\alpha \ex l}^{(\EE)} \bar{\sigma}^{(\EE)} \bar{\rho}_\ex^{(\EE)}  \bar{a}_{\ex l}^{(\EE)},
\label{approx_Pl}
\end{equation}\end{linenomath}
with the single and double averages appearing as:
\begin{linenomath}\begin{align}
\bar{\sigma}^{(\EE)}&\equiv \frac{1}{W_\EE}\int_\EE w(E)\,\sigma(E)\,\D E,
\label{sigma_mean}\\
\bar{\rho}_\ex^{(\EE)}&\equiv \frac{1}{W_\EE}\int_\EE w(E)\,\rho(\ex,E)\,\D E,
\label{rho_mean}\\
\bar{a}_{\ex l}^{(\EE)}&\equiv \frac{1}{W_\EE} \int_\EE w(E)\, a_l(\ex,E)\,\D E,
\label{axl_mean}\\
\bar{\bar{\eps}}_{\alpha \ex l}^{(\EE)}&\equiv \frac{1}{2W_\EE} \int_\EE w(E)\,\D E \int_ {-1}^1 \eps_\alpha(\ex,E,\x)\,P_l(\x)\,\D\x.
\label{eps_xl_mean}
\end{align}\end{linenomath}
In analogy to eq.~(\ref{alpha}) we introduce another arbitrary bijective mapping:
\begin{linenomath}\begin{equation}
(\ex,l)\mapsto\beta,
\label{gamma}
\end{equation}\end{linenomath}
never having to be explicitly constructed, but allowing for a unique-index labeling. In that, $\beta$ spans the range of all free parameters, i.e. the total number of coefficients $\bar{a}_{\ex l}^{(\EE)}$: \mbox{$\beta=1,...,\mathcal{P}_\EE$}. As it holds: \mbox{$\mathcal{P}_\EE=\sum_{\ex=0}^{\Xmax_\EE}[\Lmax_\EE(\ex)+1]$}, from eq.~(\ref{constraint}) we have the following constraint upon the distribution of Legendre coefficients among the relevant exited states:
\begin{linenomath}\begin{equation}
1+\Xmax_\EE+\sum_{\ex=0}^{\Xmax_\EE}\Lmax_\EE(\ex)\le\mathcal{R}_\EE.
\label{L_constraint}
\end{equation}\end{linenomath}
Equation~(\ref{recast_Pl}) is now recast as:
\begin{linenomath}\begin{align}
\begin{split}
N_\alpha^{(\EE)}&\approx 2W_\EE \sum_{\ex=0}^{\Xmax_\EE}\sum_{l=0}^{\Lmax_\EE(\ex)} \bar{\bar{\eps}}_{\alpha \ex l}^{(\EE)} \bar{\sigma}^{(\EE)} \bar{\rho}_\ex^{(\EE)} \bar{a}_{\ex l}^{(\EE)} = 2W_\EE \sum_{\beta=1}^{\mathcal{P}_\EE} \bar{\bar{\eps}}_{\alpha \beta}^{(\EE)} \bar{\sigma}^{(\EE)} \bar{\rho}_\beta^{(\EE)} \bar{a}_{\beta}^{(\EE)},
\end{split}
\end{align}\end{linenomath}
having thus been brought into the matrix form from eq.~(\ref{vectorized}), with the appropriate definitions:
\begin{linenomath}\begin{align}
\big[\Emx\big]_{\alpha\beta}&\equiv 2W_\EE \bar{\bar{\eps}}_{\alpha \beta}^{(\EE)},
\label{Emx_xl}\\
\prt_{\beta}^{(\EE)}&\equiv \bar{\sigma}^{(\EE)} \bar{\rho}_\beta^{(\EE)} \bar{a}_{\beta}^{(\EE)}.
\label{P_xl}
\end{align}\end{linenomath}
The entire solution $\vec{\prt}^{\,(\EE)}$ and the corresponding uncertainties are now easily found from Eqs.~(\ref{lsq_mean}) and (\ref{lsq_uncer}).

Applying the normalization condition \mbox{$\int_{-1}^1 A(\ex,E,\x) \D\x=1$} to eq.~(\ref{A_decomp}), we find that:
\begin{linenomath}\begin{equation}
a_0(\ex,E)=1/2 \quad\Rightarrow\quad \bar{a}_{\ex 0}^{(\EE)}=1/2,
\label{a0_fix}
\end{equation}\end{linenomath}
i.e. all the $0^\mathrm{th}$ terms are fixed and carry the entire angular distribution norm. Plugging this result into eq.~(\ref{P_xl}): \mbox{$\prt_{\ex 0}^{(\EE)}= \bar{\sigma}^{(\EE)} \bar{\rho}_\ex^{(\EE)}/2$} and combining it with the normalization condition \mbox{$\sum_{\ex=0}^{\Xmax_\EE} \bar{\rho}_\ex^{(\EE)}=1$}, we find:
\begin{linenomath}\begin{equation}
\bar{\sigma}^{(\EE)}=2\sum_{\ex=0}^{\Xmax_\EE}\prt_{\ex 0}^{(\EE)}.
\label{sigma_win}
\end{equation}\end{linenomath}
The next step consists of identifying the branching ratios as:
\begin{linenomath}\begin{equation}
\bar{\rho}_\ex^{(\EE)}=\frac{2\prt_{\ex 0}^{(\EE)}}{\bar{\sigma}^{(\EE)}}=\frac{\prt_{\ex 0}^{(\EE)}}{\sum_{\ey=0}^{\Xmax_\EE}\prt_{\ey 0}^{(\EE)}},
\label{rho}
\end{equation}\end{linenomath}
culminating in the separation of the angular coefficients:
\begin{linenomath}\begin{equation}
\bar{a}_{\ex l}^{(\EE)}=\frac{\prt_{\ex l}^{(\EE)}}{\bar{\sigma}^{(\EE)}\bar{\rho}_\ex^{(\EE)}}=\frac{1}{2}\frac{\prt_{\ex l}^{(\EE)}}{\prt_{\ex 0}^{(\EE)}}.
\label{axl}
\end{equation}\end{linenomath}
The uncertainties $\delta\bar{\sigma}^{(\EE)}$, $\delta\bar{\rho}_\ex^{(\EE)}$ and $\delta\bar{a}_{\ex l}^{(\EE)}$ follow directly from eq.~(\ref{uncer}), according to the full covariance matrix $\Wmx$ from eq.~(\ref{lsq_uncer}) .

When the total number of the relevant excited states becomes so large that the total number $\mathcal{P}_\EE$ of required parameters becomes comparable to the total number $\mathcal{R}_\EE$ of available data points, and/or when these points are affected by large uncertainties, the coefficients $\bar{a}_{\ex l}^{(\EE)}$ exhibit substantial uncertainties themselves and the contributions from the particular excited states can not be reliably separated. In this case one may attempt to reconstruct the "global" partial wave coefficients averaged over the excited states:
\begin{linenomath}\begin{equation}
\bar{\bar{a}}_{l}^{(\EE)}=\sum_{\ex=0}^{\Xmax_\EE}  \bar{\rho}_\ex^{(\EE)}\bar{a}_{\ex l}^{(\EE)} f_{\Lmax_\EE(\ex)-l}=\frac{\sum_{\ex=0}^{\Xmax_\EE}   \prt_{\ex l}^{(\EE)} f_{\Lmax_\EE(\ex)-l}}{2\sum_{\ey=0}^{\Xmax_\EE}\prt_{\ey 0}^{(\EE)}},
\end{equation}\end{linenomath}
hoping for a manageable uncertainty in the total contribution to a given partial wave. The factors $f_\ell$,defined as:
\begin{linenomath}\begin{equation}
f_\ell\equiv\left\{\begin{array}{lll}
0&\mathrm{if}&\ell<0\\
1&\mathrm{if}&\ell\ge0
\end{array}\right.
\end{equation}\end{linenomath}
simply take into account whether a given partial wave was adopted for a given excited state. As all the $0^\mathrm{th}$ terms are fixed by eq.~(\ref{a0_fix}), we immediately have \mbox{$\bar{\bar{a}}_0^{(\EE)}=1/2$} and \mbox{$\delta\bar{\bar{a}}_0^{(\EE)}=0$}.

\section{Method implementation}
\label{demonstration}

We illustrate the implementation of the method by applying it to the $^{12}$C($n,p$) data artificially generated by the Geant4 simulations. We consider here the data from 1~MeV wide energy range \mbox{$\EE= [19.5\,\mathrm{MeV},20.5\,\mathrm{MeV}]$}, approximately where this reaction's cross section is expected to reach its maximum. The branching ratios and the angular distributions for each relevant excited state were arbitrarily constructed.

\subsection{Selecting the excited states}

\begin{figure}[t!]
\centering
\includegraphics[width=\width\linewidth,keepaspectratio]{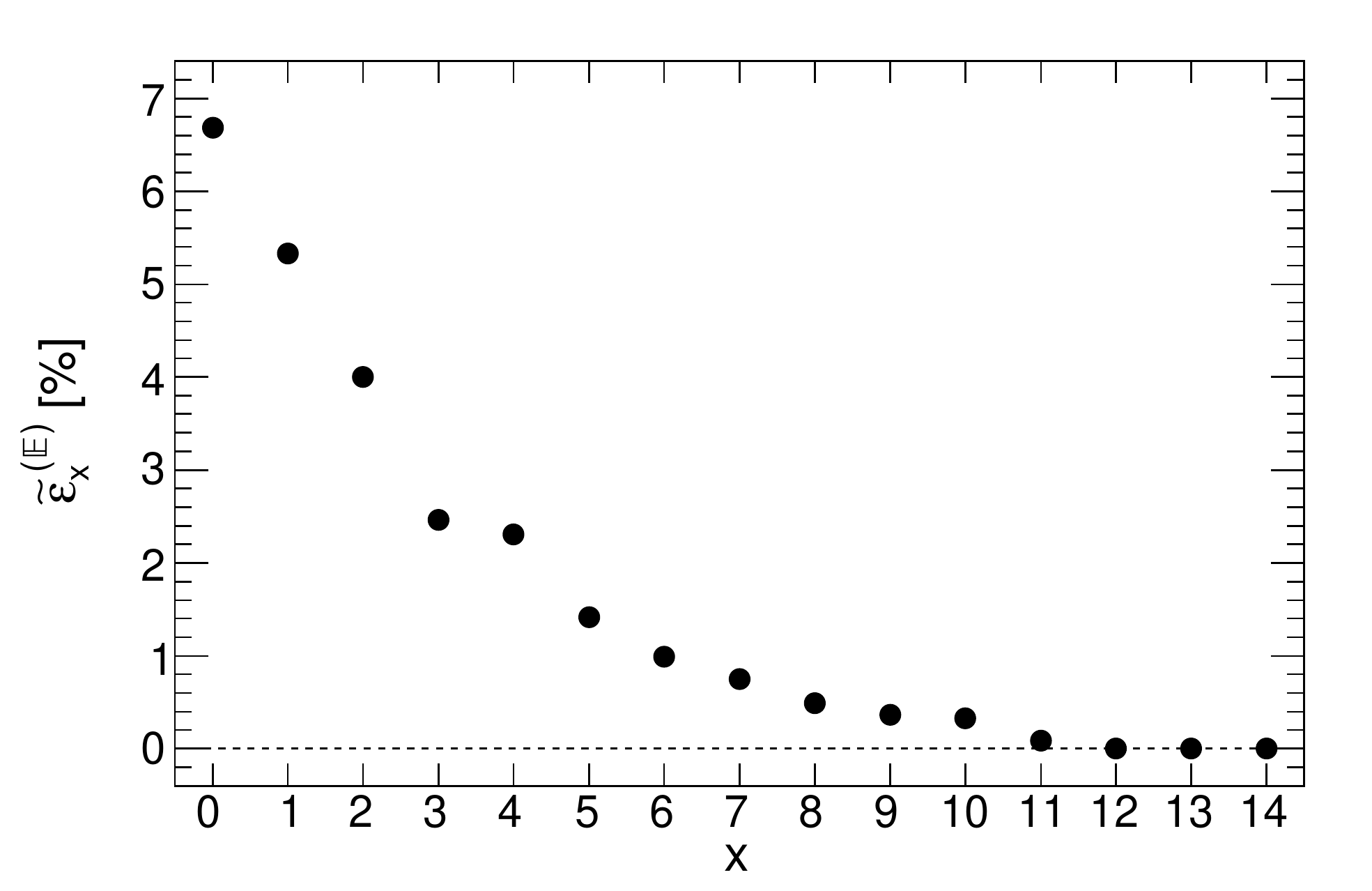}
\caption{Figure of merit: estimated probabilities for the coincidental detection of protons from the $^{12}$C($n,p$) reaction by any pair of $\Delta \mathrm{E}$-$\mathrm{E}$ strips, dependent on the excited state $\ex$ that the daughter nucleus $^{12}$B was left in. The considered neutron energy range is \mbox{$\EE= [19.5\,\mathrm{MeV},20.5\,\mathrm{MeV}]$}.}
\label{fig4}
\end{figure}

The excited states contributing to the reaction within the given neutron energy range $\EE$ need to be clearly identified, as the method requires them to be known in advance. For the $^{12}$C($n,p$) reaction, there are total of 15 states in the $^{12}$B daughter nucleus with the energy threshold $E_\mathrm{thr}$ below the upper limit of the considered neutron energy range (\mbox{$E_\mathrm{thr}<20.5\,~\mathrm{MeV}$}) \cite{b12_states}. Their excitation energies together with the corresponding $Q$-values and the energy thresholds in the laboratory frame are listed in table~\ref{tab1}. While all these states contribute to the reaction, not all of them necessarily contribute to the totality of the detected counts, especially those very close to the reaction threshold. The reason is threefold: (1)~the very low reaction cross section close to the threshold; (2)~the pronounced forward boost of the produced protons in the laboratory frame, making them miss most of the detection setup; (3)~the low proton production energy causing them to be stopped by the sample itself, never reaching the detectors at all. Therefore, it needs to be estimated in advance which states may be excluded from the analysis of the experimental data. As the exact evaluation of the expected amount of the detected counts from each state is, of course, impossible without the prior knowledge of the partial cross sections for each separate state (their branching ratios and angular distributions), one needs to rely on some reasonable estimate. One such useful figure of merit is the approximate probability estimator $\tilde{\eps}_\ex^{(\EE)}$ for the coincidental detection by any pair of the $\Delta \mathrm{E}$-$\mathrm{E}$ strips:
\begin{linenomath}\begin{equation}
\tilde{\eps}_\ex^{(\EE)}\equiv \frac{1}{2W_\EE}\sum_\alpha\int_\EE w(E)\D E \int_{-1}^1 \eps_\alpha(\ex,E,\x) \D\x,
\end{equation}\end{linenomath}
constructed by assuming --- in the absence of any prior information --- the isotropic angular distribution of protons in the center of mass frame: \mbox{$A(\ex,E,\x)\approx 1/2$}, and applying the same energy deposition cuts as are to be applied to the experimental data. Figure~\ref{fig4} shows thus obtained probability estimates for the relevant $^{12}$B states. Although the portion $\N_\ex^{(\EE)}$ of the produced protons still remains entirely unknown, the observed decrease in $\tilde{\eps}_\ex^{(\EE)}$ together with the expected decrease in $\N_\ex^{(\EE)}$ for the higher states allows one to make informed estimates about the relevance of the expected partial contributions $N_\ex^{(\EE)}$ to the detected counts: \mbox{$N_\ex^{(\EE)}\approx \tilde{\eps}_\ex^{(\EE)} \N_\ex^{(\EE)}$}. From these considerations applied to figure~\ref{fig4} we elect to include only the first 11 states (up to the $10^\mathrm{th}$ excited one, i.e. \mbox{$\Xmax_\EE=10$}) for further analysis. The artificial data to be analyzed were, of course, simulated by including all 15 states with the energy thresholds below the upper limit of the considered neutron energy range.

\begin{table}[t!]
\caption{States in the $^{12}$B nucleus relevant for the selected demonstration example. The table lists their excitation energies $E_\ex$ \cite{b12_states}, the corresponding $Q$-values and the energy thresholds $E_\mathrm{thr}$ for the $^{12}$C($n,p$) reaction in the laboratory frame.}
\label{tab1}
\centering
\begin{tabular}{|cccc|}
\hline
$\boldsymbol{\ex}$ & $\boldsymbol{E_\ex \; [\mathrm{MeV}]}$ & $\boldsymbol{Q \; [\mathrm{MeV}]}$ & $\boldsymbol{E_\mathrm{thr} \; [\mathrm{MeV}]}$\\
\hline
0 & 0.00 & 12.59 & 13.65 \\
1 & 0.95 & 13.54 & 14.68 \\
2 & 1.67 & 14.26 & 15.46 \\
3 & 2.62 & 15.21 & 16.49 \\
4 & 2.73 & 15.31 & 16.60 \\
5 & 3.39 & 15.98 & 17.32 \\
6 & 3.76 & 16.35 & 17.72 \\
7 & 4.00 & 16.59 & 17.98 \\
8 & 4.30 & 16.89 & 18.31 \\
9 & 4.46 & 17.05 & 18.48 \\
10 & 4.52 & 17.11 & 18.55 \\
11 & 4.99 & 17.56 & 19.05 \\
12 & 5.61 & 18.20 & 19.73 \\
13 & 5.73 & 18.31 & 19.85 \\
14 & 6.00 & 18.59 & 20.15 \\
\hline
\end{tabular}
\end{table}

It must be pointed out that this exclusion of higher states from the analysis may, in principle, affect the cross section normalization, as the branching ratios of the excluded states become unobtainable. However, as already discussed, the cross sections around the energy thresholds for these states are expected to be negligible and so is their impact upon the total reaction cross section. Still, if there were reasonable indications to the contrary, one should be aware that the reconstructed cross section $\bar{\sigma}^{(\EE)}$ is only partially contributed by those states that were kept for the analysis.

\subsection{Varying the amount of partial waves}
\label{partials}

The highest wave numbers $\Lmax_\EE(\ex)$ for each excited state are evidently the method's adjustable parameters. For the total of $\mathcal{R}_\EE$ relevant pairs of strips from eq.~(\ref{constraint}), there is a total of $\binom{\mathcal{R}_\EE}{\Xmax_\EE+1}$ selections of $\Lmax_\EE(\ex)$ satisfying the constraint from eq.~(\ref{L_constraint}), with $\binom{\,\cdot\,}{\,\cdot\,}$ denoting a binomial factor. For \mbox{$\mathcal{R}_\EE=60$}, as used later, and \mbox{$\Xmax_\EE=10$} this amounts to approximately \mbox{$3.4\times10^{11}$} combinations. If we were to impose some maximum admissible wave number $\maxL$ that may be assigned to any particular state --- implying, for purpose of these simple estimates, that the selection of $\maxL$ itself must be such that \mbox{$(\maxL+1)(\Xmax_\EE+1)\le\mathcal{R}_\EE$}, in order for each of \mbox{$\Xmax_\EE+1$} states to be allowed \mbox{$\maxL+1$} waves --- then the number of possible selections for $\Lmax_\EE(\ex)$ reduces to \mbox{$(\maxL+1)^{\Xmax_\EE+1}$}. For example, the maximum value $\maxL=4$ compatible with \mbox{$\mathcal{R}_\EE=60$} and \mbox{$\Xmax_\EE=10$} yields approximately \mbox{$4.9\times10^7$} combinations. However, the following physical argument helps us in reducing the number of possible combinations even further, by keeping only the physically sensible selections of $\Lmax_\EE(\ex)$. We consider that close to the reaction threshold the nuclear reactions are expected to be isotropic (in the center-of-mass frame), while the anisotropy is expected to appear (and possibly intensify) with increasing incident particle energy. This suggests that the higher excited states --- characterized by a higher threshold --- should not be assigned more partial waves than the lower states, i.e.:
\begin{linenomath}\begin{equation}
\Lmax_\EE(\ex_1)\ge\Lmax_\EE(\ex_2) \quad\mathrm{for}\quad \ex_1<\ex_2.
\label{}
\end{equation}\end{linenomath}
For the maximum admissible wave number $\maxL$, the number of combinations consistent with this constraint is now reduced to $\binom{\maxL+\Xmax_\EE+1}{\Xmax_\EE+1}$. For example, the maximal value $\maxL=4$ compatible with \mbox{$\mathcal{R}_\EE=60$} and \mbox{$\Xmax_\EE=10$} leaves the total of 1365 combinations. All we need now is the algorithm for constructing such combinations. For the maximum wave number $\maxL$ to be assigned to any state, the particular combination of nonincreasing $\Lmax_\EE(\ex)$ values may be uniquely defined by the set of $\maxL$ states $\Lambda_\ell$ (\mbox{$\ell=1,\ldots,\maxL$}) at which the maximum wave number $\Lmax_\EE(\ex)$ increases by 1. In other words, $\Lambda_\ell$ form a set of states such that \mbox{$\Lmax_\EE(\ex)=\ell$} ends at \mbox{$\ex=\Lambda_\ell$}, i.e.:
\begin{linenomath}\begin{equation}
\Lmax_\EE(\ex)=\ell \quad\mathrm{for}\quad \Lambda_{\ell+1}<\ex\le\Lambda_\ell,
\label{max_wave}
\end{equation}\end{linenomath}
with additional fixed boundaries \mbox{$\Lambda_0=\Xmax_\EE$} and \mbox{$\Lambda_{\maxL+1}=-1$}, defined for the convenience of the implementation. The algorithm now reduces to generating all combinations ($\maxL$-tuples) of $\Lambda_\ell$ such that:
\begin{linenomath}\begin{equation}
\Lambda_{\ell+1}\le\Lambda_\ell \quad\mathrm{with}\quad \Lambda_\ell\in\big\{-1,\ldots,\Xmax_\EE\big\} \quad\mathrm{for}\quad \ell=1,\ldots,\maxL.
\label{vary}
\end{equation}\end{linenomath}
It is easy to confirm that if \mbox{$\Lambda_\ell=-1$} for all $\ell$, then \mbox{$\Lmax_\EE(\ex)=0$} for all $\ex$, i.e. all the states are assigned only the  isotropic component. The other extreme is \mbox{$\Lambda_\ell=\Xmax_\EE$} for all $\ell$, meaning that \mbox{$\Lmax_\EE(\ex)=\maxL$} for all $\ex$, i.e. all the states are assigned the maximum allowed number of partial waves.

\subsection{Optimizing the model parameters}

The obvious question now is how to select an optimal combination of the wave numbers $\Lmax_\EE(\ex)$. We propose here a simple --- and by no means unique --- selection principle. As the variations in $\Lmax_\EE(\ex)$ directly affect the number of the model parameters: \mbox{$\mathcal{P}_\EE=\sum_{\ex=0}^{\Xmax_\EE}[\Lmax_\EE(\ex)+1]$}, the reduced chi-squared estimator $\mathrm{X}^2$ lends itself easily to a quick and efficient evaluation of the \textit{goodness} of the fit:
\begin{linenomath}\begin{align}
\begin{split}
\mathrm{X}^2=\frac{\big(\vec{N}^{(\EE)}-\Emx \vec{\prt}^{\,(\EE)}\big)^\top \Vmx^{-1} \big(\vec{N}^{(\EE)}-\Emx \vec{\prt}^{\,(\EE)}\big)}{\mathcal{R}_\EE-\mathcal{P}_\EE}\simeq \frac{1}{\mathcal{R}_\EE-\mathcal{P}_\EE}\sum_{\alpha=1}^{\mathcal{R}_\EE}\frac{\big(N_\alpha^{(\EE)}-\sum_{\beta=1}^{\mathcal{P}_\EE} \big[\Emx\big]_{\alpha\beta} \prt_\beta^{\,(\EE)} \big)^2}{\big(\delta N_\alpha^{(\EE)}\big)^2}.
\end{split}
\label{chi2}
\end{align}\end{linenomath}
The rightmost expression holds when the covariance matrix $\Vmx$ of the input data is diagonal, i.e. when the correlations between the components of $\vec{N}^{(\EE)}$ are negligible. As opposed to the goodness of fit --- which will for large $\mathcal{R}_\EE$ systematically improve by increasing the number of partial waves, as long as $\mathcal{P}_\EE$ does not closely approach $\mathcal{R}_\EE$ --- the \textit{reliability} of the fit, reflected through the uncertainties in the reconstructed $\vec{\prt}^{\,(\EE)}$, rapidly degrades with increasing number of parameters. For estimating this reliability we propose a simple calculation of the uncertainty $\delta\mathrm{X}^2$ in the chi-squared value from eq.~(\ref{chi2}) by means of eq.~(\ref{uncer}), since $\mathrm{X}^2$ is sensitive to \textit{all} the fitted parameters --- unlike, for example, the reconstructed cross section $\bar{\sigma}^{(\EE)}$ from eq.~(\ref{sigma_win}). In the context of our problem the minimization of $\mathrm{X}^2$ and its uncertainty $\delta\mathrm{X}^2$ seem to be opposing objectives. Therefore, we propose to minimize their product $\mathrm{X}^2\delta\mathrm{X}^2$ as the simplest estimator that should at its minimum provide the optimal tradeof between the goodness and the reliability of the fit.

There are additional issues to consider. For the number of partial waves too inadequate for a given set of the experimental data, some of the branching ratios $\bar{\rho}_\ex^{(\EE)}$ from eq.~(\ref{rho}) may turn out to be negative or greater than unity --- a clear signature of the badness of the fit, going beyond the particular values of $\mathrm{X}^2$. These fits should be immediately rejected as physically unsound, i.e. disqualified from any kind of optimization procedure, be it the minimization of $\mathrm{X}^2\delta\mathrm{X}^2$ or some alternate technique.

Yet another quality control mechanism consists of checking if the reconstructed angular distributions for each angular state:
\begin{linenomath}\begin{align}
\begin{split}
\bar{A}_\ex^{(\EE)}(\x)&\equiv \sum_{l=0}^{\Lmax_\EE(\ex)} \bar{a}_{\ex l}^{(\EE)} P_l(\x)= \frac{1}{2}\sum_{l=0}^{\Lmax_\EE(\ex)} \frac{\prt_{\ex l}^{(\EE)}}{\prt_{\ex 0}^{(\EE)}}P_l(\x)
\end{split}
\end{align}\end{linenomath}
become negative at any point. If so, such fits may also be immediately rejected, regardless of their goodness. One should be wary, however, in making such rejections when there are prior indications that some states may indeed feature the very low branching ratios or highly anisotropic angular distributions that locally come close to zero. In this case any statistical fluctuation in the input data may easily drive the reconstructed results toward the negative values, while the results do remain reasonably reliable representations of the true reaction parameters. It should be noted that the reconstructed branching ratios may discard \textit{all} fits as unphysical, if every combination of wave numbers $\Lmax_\EE(\ex)$ produces at least one negative $\bar{\rho}_\ex^{(\EE)}$. On the other hand, the isotropic angular distributions will always pass the negativity test, so that the fully isotropic fit (\mbox{$\Lmax_\EE(\ex)=0$} for all~$\ex$) will always be accepted, based on the positivity of the angular distributions themselves.

Prior to calculating $\mathrm{X}^2$, $\delta\mathrm{X}^2$ or any consequent quantity to be used in judging the suitability of the fitted result, one may also consider manually eliminating from the set of fitted parameters those $\prt_{\ex l}^{(\EE)}$ that, according to eq.~(\ref{axl}), yield the angular coefficients too small (\mbox{$|\bar{a}_{\ex l}^{(\EE)}|\ll1$}) or unreasonably large (\mbox{$|\bar{a}_{\ex l}^{(\EE)}|\gg1$}) in magnitude. For the associated $\beta$, this is most easily done by setting \mbox{$\prt_\beta^{(\EE)}=0$} and \mbox{$[\Wmx]_{\beta\beta'}=[\Wmx]_{\beta'\beta}=0$} for all $\beta'$ within the covariance matrix from eq.~(\ref{lsq_uncer}). This procedure helps in regularizing the fit, as the exceedingly small \mbox{$|\bar{a}_{\ex l}^{(\EE)}|$} are commonly the sporadic results caused by the finite precision data, while the distinctly large \mbox{$|\bar{a}_{\ex l}^{(\EE)}|$} are expected to appear as the consequence of overfitting the statistical fluctuations in the input data. One should, of course, be prepared for the closer inspection and the critical evaluation of the results if the optimal set of parameters happens to be precisely thus manipulated set. However, what is expected from this procedure is the artificially induced increase in the fit suitability estimator $\mathrm{X}^2\delta\mathrm{X}^2$, such that some alternative set of parameters takes precedence as the optimal one.

In summary, we propose to identify the optimal combination of the maximum wave numbers $\Lmax_\EE(\ex)$ by minimizing the product $\mathrm{X}^2\delta\mathrm{X}^2$ --- or any such estimator balancing between the goodness and the reliability of the fit --- while taking into account the physical soundness of the results, whether by immediately rejecting those physically inadmissible or by appropriately penalizing them during the optimization procedure.

\subsection{Method investigation: $^{12}$C(\emph{n,p}) data}

We now test the method on a particularly challenging example of the artificially generated $^{12}$C($n,p$) data, as means of appraising its applicability to the experimental data from n\_TOF. The simulated dataset --- the set of counts $N_\alpha^{(\EE)}$ detected by a particular pair of strips --- was obtained from the same Geant4 simulations as used for obtaining the coincidental detection probabilities, i.e. the central design matrix $\Emx$. The neutron energies were sampled within the 1~MeV wide interval \mbox{$\EE= [19.5\,\mathrm{MeV},20.5\,\mathrm{MeV}]$}, all 15 states from table~\ref{tab1} were used in constructing the dataset, while only the first 11 states from figure~\ref{fig4} were considered for the reconstruction. For the buildup of the test counts an arbitrarily constructed branching ratios $\rho(\ex,E)$ for each of the 15 states were used (represented by later figure~\ref{fig7}), together with the angular distributions $A(\ex,E,\x)$ arbitrarily constructed for each state, which were all designed from the three lowest Legendre polynomials ($P_0$, $P_1$, $P_2$).

\begin{figure}[t!]
\centering
\includegraphics[width=\width\linewidth,keepaspectratio]{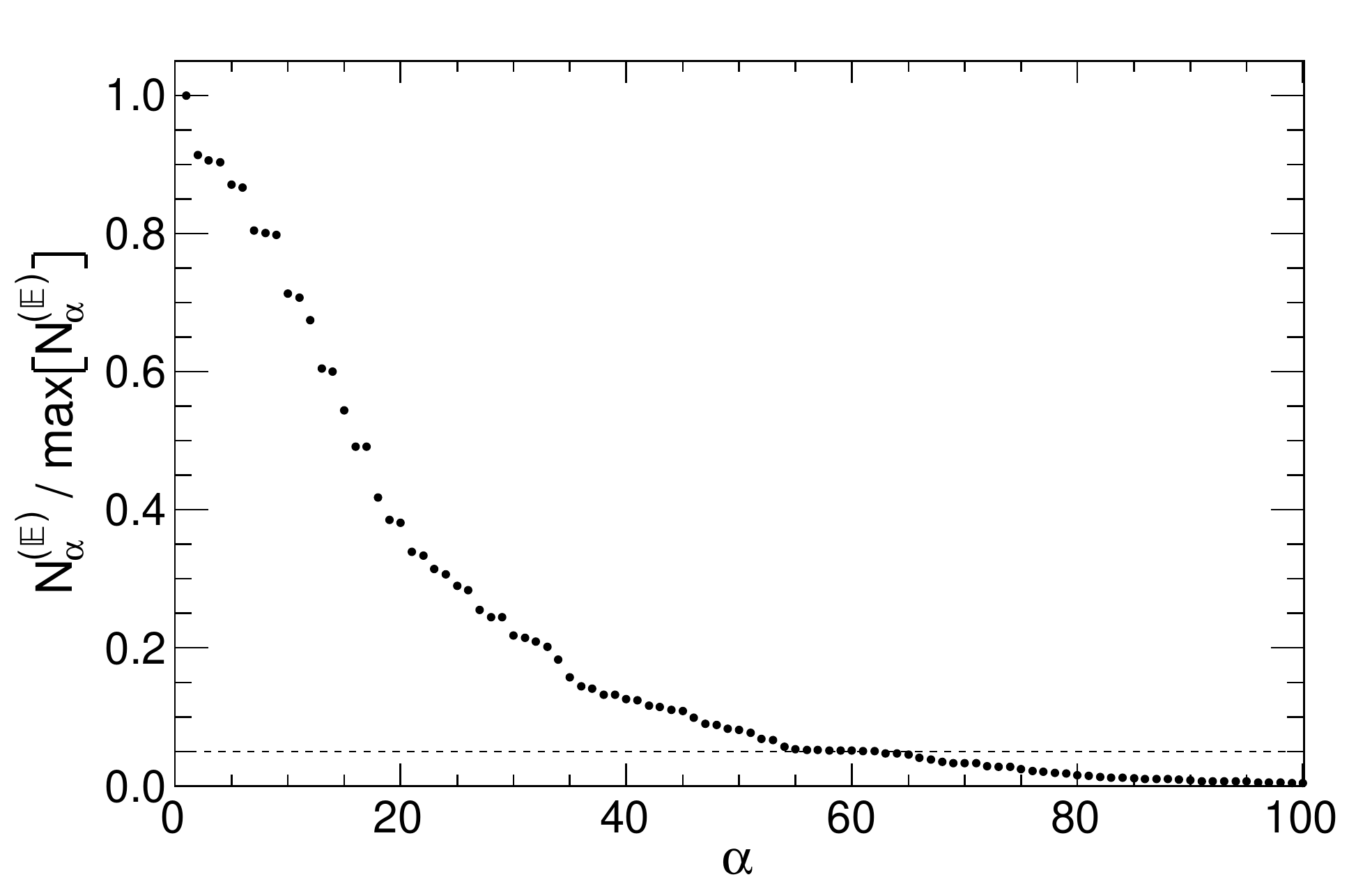}
\caption{Artificial set of the coincidentally detected counts obtained from an exceedingly large dataset generated by Geant4 simulations, virtually unaffected by the statistical fluctuations. The numbers of counts are ordered by their magnitude and scaled relative to their maximum value (from the most efficient \mbox{$\Delta \mathrm{E}$-$\mathrm{E}$} pair of strips). The values for the analysis are constructed by first scaling these counts to a desired level of statistics and then generating the appropriate Poissonian fluctuations. Only the counts above 5\% of the maximum value (the dashed threshold) are kept for the analysis.}
\label{fig5}
\end{figure}

Figure~\ref{fig5} shows the relevant set of coincidental counts recorded by different $\Delta \mathrm{E}$-$\mathrm{E}$ pairs of strips, ordered by magnitude. While there are (16~$\mathrm{E}$-strips)$\times$(16~$\Delta\mathrm{E}$-strips )$\times$(2~telescopes)~=~512 possible pairs of strips in the used SITE configuration from figure~\ref{fig0_b}, one can see from figure~\ref{fig5} that only the tenth of those are characterized by a sufficient coincidental detection probability to be considered for analysis. It should be noted that the counts from figure~\ref{fig5} were constructed from an exceedingly large dataset, featuring the negligible statistical fluctuations. In order to easily generate the statistical variations in the dataset to be taken for analysis, we first scale these counts to a desired level of statistics (thus constructing their statistically expected values) and then generate the appropriate Poissonian fluctuations. For purposes of this demonstration we keep only those coincidental counts $N_\alpha^{(\EE)}$ that are higher than 5\% of the maximum value (the dashed threshold from figure~\ref{fig5}). Depending on a particular realization of the Poissonian fluctuations, around \mbox{$\mathcal{R}_\EE=60$} relevant pairs of strips meet this condition. The statistical uncertainties are then assigned to these counts by setting the diagonal elements of the input covariance matrix $\Vmx$ from Eqs.~(\ref{lsq_mean}) and (\ref{lsq_uncer}) to: \mbox{$[\Vmx]_{\alpha\alpha}=N_\alpha^{(\EE)}$}.

In order to vary the maximum wave numbers $\Lmax_\EE(\ex)$ for each excited state we follow the procedure from section~\ref{partials}, adopting the maximum supported value \mbox{$\maxL=4$}. We choose the number of counts from the most efficient pair of strips to be: \mbox{$\max\big[N_\alpha^{(\EE)}\big]=10^6$}, making the total number of counts detected across all kept pairs: $\sum_{\alpha=1}^{\mathcal{R}_\EE}N_\alpha^{(\EE)}=2\times10^7$. The reason behind this selection is rather simple and carries the critical repercussions for the analysis of the experimental data from n\_TOF: at lower statistics basically all the fits are discarded due to the appearance of the negative branching ratios. In other words, for almost all generated instances of Poissonian fluctuations all the fits (for any combination of state boundaries $\Lambda_\ell$) produce at least one negative $\bar{\rho}_\ex^{(\EE)}$. One must be careful at this point not to confuse \mbox{$\max\big[N_\alpha^{(\EE)}\big]=10^6$} with some minimum intrinsic level of reliable statistics. Instead, it reflects an amount of excited states at play: a high number of states naturally requires high statistics if they were to be reliably disentangled one from the other.

We now appraise the method based on the accuracy and uncertainty of the reconstructed parameters. For a condensed demonstration of the results on the reconstructed angular distributions, we we use the \textit{overall distribution} $\mathcal{A}_\EE(\x)$, averaged over all excited states:
\begin{linenomath}\begin{align}
\begin{split}
\mathcal{A}_\EE(\x)&=\frac{1}{W_\EE}\sum_{\ex=0}^{\Xmax_\EE^+} \int_\EE w(E)\, \rho(\ex,E)\, A(\ex,E,\x)\,\D E\\
&\simeq\sum_{\ex=0}^{\Xmax_\EE} \bar{\rho}_\ex^{(\EE)} \sum_{l=0}^{\Lmax_\EE(\ex)} \bar{a}_{\ex l}^{(\EE)} P_l(\x)=\frac{\sum_{\ex=0}^{\Xmax_\EE} \sum_{l=0}^{\Lmax_\EE(\ex)} \prt_{\ex l}^{(\EE)}P_l(\x) }{2\sum_{\ex=0}^{\Xmax_\EE}\prt_{\ex 0}^{(\EE)}}.
\end{split}
\label{A_rec}
\end{align}\end{linenomath}
The reference distribution stems from the arbitrarily constructed distributions $A(\ex,E,\x)$ for each of the 15 states contributing to the reaction (\mbox{$\Xmax_\EE^+=14$}; see table~\ref{tab1}). The reconstructed distribution, as denoted by $\simeq$, is contributed by the reduced number of states taken for the analysis (\mbox{$\Xmax_\EE=10$}). After applying the method to the different realizations of Poissonian fluctuations, our conclusions are rather simple and straightforward. The fits yielding an unphysical set of branching ratios also grossly misidentify the overall angular distribution $\mathcal{A}_\EE(\x)$ and, in general, the reconstructed cross section $\bar{\sigma}^{(\EE)}$, reflecting the absolute normalization of the data. As such, they should indeed be immediately rejected. Among the physically admissible fits (if there are any at all) the ones identified as optimal do seem to reasonably reconstruct both the overall angular distribution and the cross section, at least under the level of statistics adopted here out of necessity. However, the set of reconstructed branching ratios themselves most often seems to be unrepresentative of the true results, as illustrated by a typical example from figure~\ref{fig7}. The example from figure~\ref{fig7} also shows that their uncertainties may also be grossly underestimated and unrepresentative of their error. Therefore, the reconstructed branching ratios should be taken with maximum caution.


\begin{figure}[t!]
\centering
\includegraphics[width=\width\linewidth,keepaspectratio]{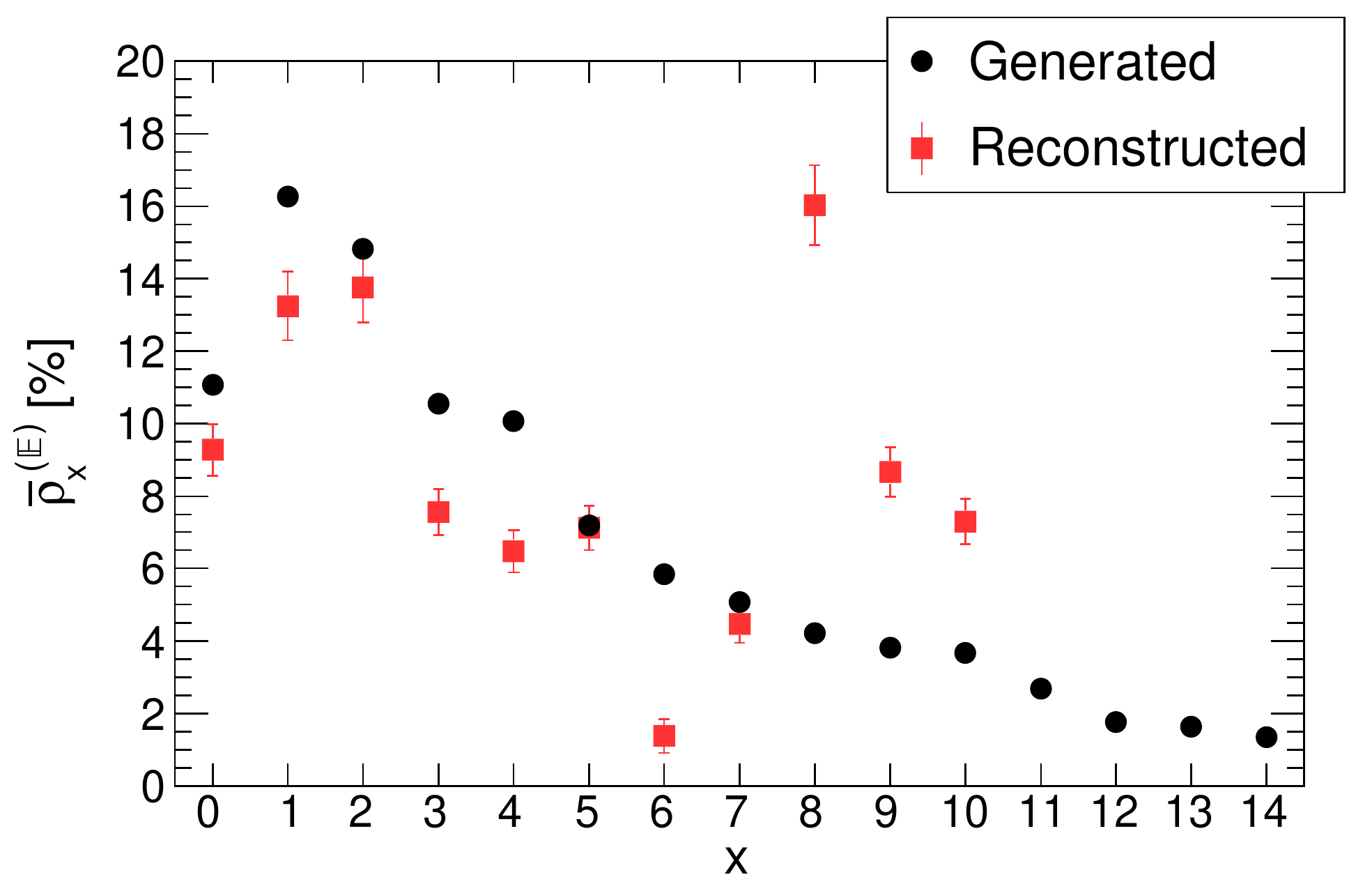}
\caption{Typical example of the reconstructed set of branching ratios, recovered by an optimal set of wave numbers assigned to each excited state. Only the first 11 states were considered for the reconstruction, as the rest of them hardly contribute to the detected counts or not at all.}
\label{fig7}
\end{figure}

\begin{figure}[t!]
\centering
\includegraphics[width=\width\linewidth,keepaspectratio]{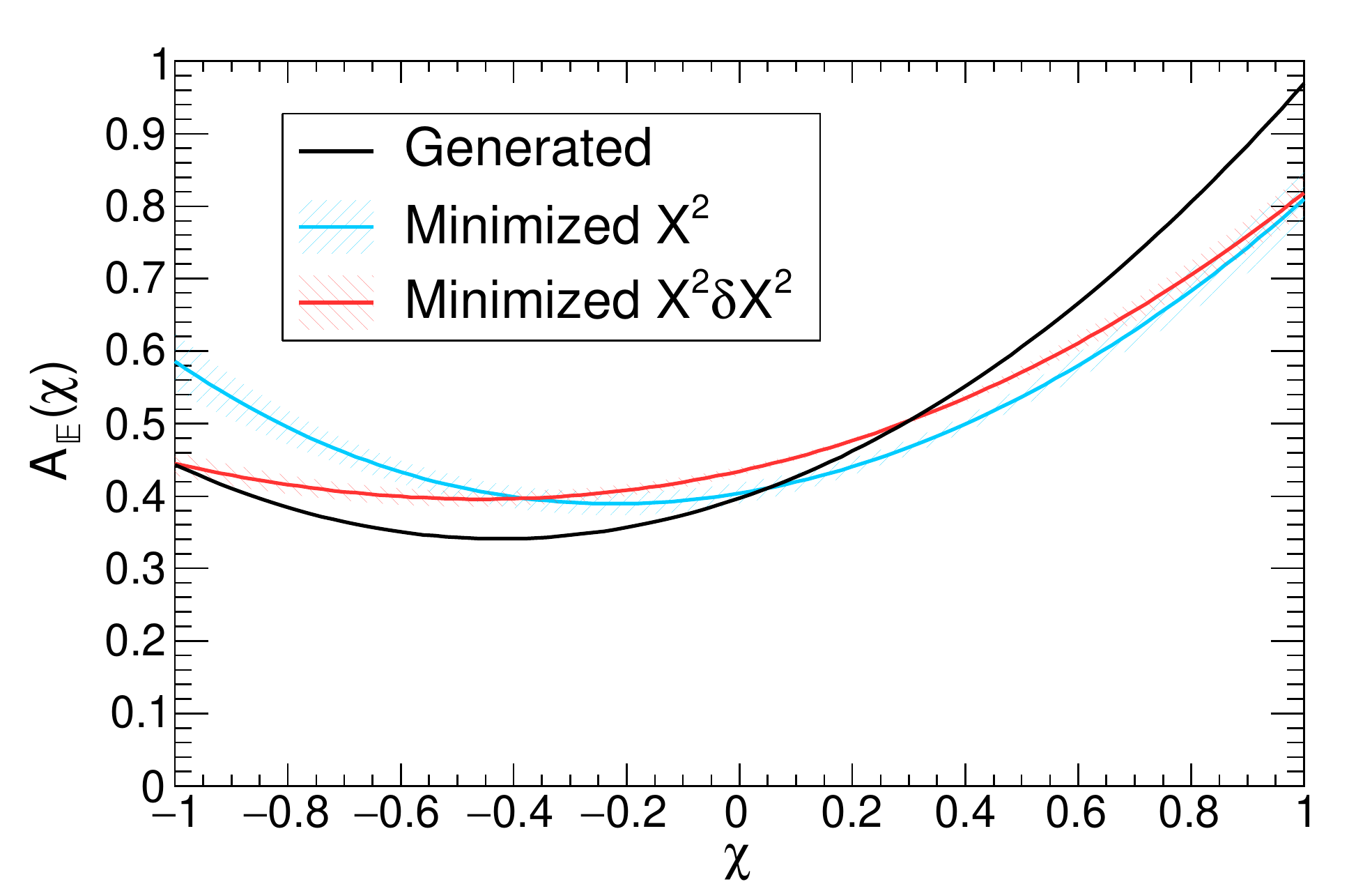}
\caption{Overall angular distribution recovered from an optimal set of wave numbers for each excited state, obtained by minimizing either $\mathrm{X}^2$ (the goodness of the fit) or $\mathrm{X}^2\delta\mathrm{X}^2$ (the tradeoff between the goodness and the reliability).}
\label{fig8}
\end{figure}

At the adopted level of statistics most often there seems to be little difference in the results obtained by minimizing $\mathrm{X}^2$ or the proposed product $\mathrm{X}^2\delta\mathrm{X}^2$, as the physicality of the branching ratios serves as the main discriminator of the unreliable fits. Figure~\ref{fig8} shows an example when the difference in the reconstructed overall angular distributions obtained by minimizing $\mathrm{X}^2$ or $\mathrm{X}^2\delta\mathrm{X}^2$ turns out to be appreciable. This example clearly illustrates the superiority of optimizing the tradeoff between he goodness and the reliability of the fit. The power of this procedure lies not in reducing the uncertainties \textit{per se}, but in penalizing the overfitting, i.e. in  rejecting the sporadic parameters that unnecessarily and disproportionately increase the uncertainties in all other parameters, besides introducing their own excessive ones. Indeed, while the reference angular distribution from figure~\ref{fig8} was constructed as a linear combination of the 3 lowest Legendre polynomials, the one identified by minimizing $\mathrm{X}^2$ allows for 5 of them (the maximum amount supported by \mbox{$\maxL=4$}; a clear symptom of overfitting), while the minimization of $\mathrm{X}^2\delta\mathrm{X}^2$ finds the combination of 4 partial waves as the optimal one.

Let us recall that with so many exited states at play, the physicality of the branching ratios serves as the primary discriminator of unreliable fits. For a significantly reduced number of states, this method of assessment becomes much more insensitive or even entirely unavailable in case of a single relevant, ground state. In that case the quality tradeoff between the goodness and the reliability of the fit remains the crucial, if not the only available method for identifying the optimal set of the fit parameters.


Finally, at the adopted level of statistics the relative uncertainty in the reconstructed cross section $\bar{\sigma}^{(\EE)}$ appears to be around 10\%. As the statistically expected uncertainty scales as \mbox{$\big(N_\tot^{(\EE)}\big)^{-1/2}$} with the total number $N_\tot^{(\EE)}$ of the detected counts, one can easily estimate the expected level of uncertainty at any level of statistics, provided that the available data produce any acceptable fit in the first place. Considering that the experimental n\_TOF data on the $^{12}$C($n,p$) reaction are expected to provide 4 to 5 orders of magnitude less statistics than adopted for this demonstration \cite{carbon_prop}, even if they could be fitted without all fits failing the physicality test, the uncertainty in $\bar{\sigma}^{(\EE)}$ is thus expected to be at least an order of magnitude grater than reconstructed cross section itself! Hence, the direct application of the full reconstruction method presented up to this point is ill-adjusted to these experimental data, due to the particularly unfavorable combination of the available statistics and the amount of excited states at play. This outcome should not be confused with some intrinsic shortcoming of the method itself, as there is a limit to the quality of the results that could be extracted from the data of a finite statistical precision. Fortunately, this eventuality was foreseen in advance of the experiment and the experimental setup was specially optimized so as to minimize the systematic effects due to the alternative approach to the analysis of these data. This approach consists of utilizing the reduced variant of the method, by adopting \textit{a priori} information on the branching ratios and the angular distributions from an outside source --- such as the TALYS theoretical model \cite{talys}, adjusted to the preexisting experimental data --- and aiming solely at the reconstruction of the absolute cross section $\bar{\sigma}^{(\EE)}$. This reduced variant is addressed in the following section.

\section{Method reduction}
\label{reduction}

Even when the full unfolding procedure may not be meaningfully applied due to the uncertainties in the input data limiting the usefulness of the output results, the method formalism from section~\ref{derivation} still remains relevant, as it clearly establishes the connection between the measured observables and the underlying reaction parameters. Furthermore, the coincidental detection probability of the experimental setup must be characterized --- most appropriately by means of the dedicated simulations described in appendix~\ref{app_eff} ---  regardless of the particular approach to the data analysis. Starting from eq.~(\ref{master}), one may derive any particular variant of the unfolding procedure, be it the reduction of the one from section~\ref{pw_decomposition} or even some further extension, shortly addressed in appendix~\ref{extension}. Motivated by the status of the experimental n\_TOF data on the $^{12}$C($n,p$) reaction, we consider here the adoption of \textit{a priori} information on the branching ratios and angular distributions, aiming solely at the reconstruction of the absolute cross section. Assuming that information to be available from an independent source, eq.~(\ref{master}) may be linearized as:
\begin{linenomath}\begin{equation}
\vec{N}^{(\EE)}\approx  \vec{\epsilon}^{\,(\EE)} \bar{\sigma}^{(\EE)} ,
\end{equation}\end{linenomath}
with the vector $\vec{\epsilon}^{\,(\EE)}$ (as a matrix of a reduced dimensionality) standing in place of the design matrix $\Emx$ from eq.~(\ref{vectorized}) and the single unknown $\bar{\sigma}^{(\EE)}$ replacing the previous set $\vec{\prt}^{\,(\EE)}$ of underlying reaction parameters. While the definition of $\bar{\sigma}^{(\EE)}$ stays the same as in eq.~(\ref{sigma_mean}), $\vec{\epsilon}^{\,(\EE)}$ is now defined by components as:
\begin{linenomath}\begin{equation}
\epsilon_\alpha^{\,(\EE)} \equiv \sum_{\ex=0}^{\Xmax_\EE}\int_\EE \D E \int_{-1}^1 \D\x \; w(E)\,\rho(\ex,E)\,A(\ex,E,\x)\,\eps_\alpha(\ex,E,\x),
\label{des_vec}
\end{equation}\end{linenomath}
where the branching ratios $\rho(\ex,E)$ and angular distributions $A(\ex,E,\x)$ are taken from an outside source of information. Applying Eqs.~(\ref{lsq_mean}) and (\ref{lsq_uncer}) --- while taking the covariance matrix $\Vmx$ to be diagonal and composed of the uncertainties $\delta N_\alpha^{(\EE)}$ in the detected counts: \mbox{$[\Vmx]_{\alpha\alpha}=\big(\delta N_\alpha^{(\EE)}\big)^2$} --- the final solution for the sought cross section may now be written in a rather simple closed form:
\begin{linenomath}\begin{equation}
\bar{\sigma}^{(\EE)}= \big(\delta\bar{\sigma}^{(\EE)}\big)^2 \sum_{\alpha=1}^{\mathcal{R}_\EE}\frac{\epsilon_\alpha^{\,(\EE)}N_\alpha^{(\EE)}}{\big(\delta N_\alpha^{(\EE)}\big)^2},
\end{equation}\end{linenomath}
with the associated uncertainty:
\begin{linenomath}\begin{equation}
\delta\bar{\sigma}^{(\EE)}=\left(\sum_{\alpha=1}^{\mathcal{R}_\EE}\Bigg(\frac{\epsilon_\alpha^{\,(\EE)}}{\delta N_\alpha^{(\EE)}}\Bigg)^2\right)^{-1/2}.
\end{equation}\end{linenomath}
It should be noted that this procedure still makes full use of all the experimentally available information from separate pairs of $\Delta \mathrm{E}$-$\mathrm{E}$ strips. This feature is in clear opposition with the more extreme reduction of the method, taking only the total number \mbox{$N_\tot^{(\EE)}=\sum_{\alpha=1}^{\mathcal{R}_\EE} N_\alpha^{(\EE)}$} of coincidental counts detected across an entire detection setup, in conjunction with its total detection probability \mbox{$\epsilon_\tot^{(\EE)}=\sum_{\alpha=1}^{\mathcal{R}_\EE} \epsilon_\alpha^{(\EE)}$} in order to obtain the absolute cross section overly simplistically as \mbox{$\bar{\sigma}^{(\EE)}=N_\tot^{(\EE)}/\epsilon_\tot^{(\EE)}$}, thus defeating any benefit of having used a high-end telescope --- in particular, its sophisticated dissociation into multiple strips.

Evidently, the main challenge with thus reduced method is the estimation of the systematic uncertainties brought on by the out-of-necessity adopted branching ratios and angular distributions. An indication of those uncertainties --- and a conservative one, at that --- may be obtained by adopting all the involved angular distributions as isotropic: \mbox{$A(\ex,E,\x)=1/2$}, and recalculating $\bar{\sigma}^{(\EE)}$. The difference between the externally provided and all-isotropic distributions is to be taken as representing the extreme case of the possible disparity with the true angular distributions. Another possibility is taking among the externally provided distributions only the branching ratios or the angular distributions as given, and unfolding the data with the other type of distributions unconstrained. Comparing these alternative results for $\bar{\sigma}^{(\EE)}$ allows for an informed estimate of the systematic uncertainties.

\section{Conclusions}
\label{conclusions}

A new angle resolving stripped silicon telescope (SITE) has recently been introduced at the neutron time of flight facility n\_TOF at CERN for the measurements of the neutron induced reactions with the charged particles in the exit channel. Its outstanding detection properties have already been demonstrated in the challenging measurement of the $^7$Be($n,p$) reaction, relevant for the famous Cosmological Lithium Problem. The joint energy-differential measurement of the $^{12}$C($n,p$) and $^{12}$C($n,d$) reactions has also been recently performed at n\_TOF, using the upgraded and specially optimized detector configuration consisting of the two separate silicon telescopes. As the nature of these reactions poses significant challenges for the meaningful data analysis --- being affected by the multiple excited states in the daughter nuclei and featuring the anisotropic angular distributions of the reaction products --- we have established a clear and detailed formalism behind the measured observables: the total number of the coincidental counts detected by any combination of $\Delta \mathrm{E}$-$\mathrm{E}$ pairs of silicon strips. From this formal connection we have developed and tested the unfolding procedure for the reconstruction of the underlying reaction parameters, consisting of the absolute reaction cross section, the branching ratios and the angular distributions  of the reaction products for each excited state in the daughter nucleus. We have also addressed the finer points of the method implementation, thus providing the consistent and reliable methodology for obtaining the optimal set of the output parameters. Though the method may, in principle, reconstruct all these quantities separately, its performance may be severely limited by the amount of parameters --- determined by the number of excited states and the level of anisotropy --- as well as the level of uncertainties in the input data. By testing the method on the artificially generated dataset resembling the n\_TOF measurement the of the $^{12}$C($n,p$) reaction, we have found little hope that the full unfolding procedure could be meaningfully applied to \emph{these particular} experimental data, precisely due to the highly unfavorable combination of the large number of the excited states and the reduced level of statistics expected from the experiment. This unfortunate outcome should not be misinterpreted for the inherent deficiency of the method itself, as at some point all considered reaction parameters must be fully taken into account if the experimental data are to be properly described and reliably analyzed. Precisely the clarity of the formalism behind the method allows for its many alternative variants to be developed. One of these, to be applied to the measured $^{12}$C($n,p$) and $^{12}$C($n,d$) data, is the reduced procedure relying on the independent source of information on the branching ratios and angular distributions, aiming at the reconstruction of the absolute cross section as the central reaction parameter of interest. It is worth noting that thus reduced method still takes advantage of the distribution of the detected counts across the separate $\Delta \mathrm{E}$-$\mathrm{E}$ pairs of strips, as opposed to considering only the total number of counts across all of them. Thus retained angular sensitivity opens the possibility for the estimation of the systematic effects due to the adopted outside information (branching ratios and/or angular distributions), allowing for the informed assessment of the systematic uncertainties in the final results.


\appendix

\section{Detection probability simulations}
\label{app_eff}

We describe the detection probability simulations and the use of the simulated data in the construction of the design matrix $\Emx$ from eq.~(\ref{Emx_xl}). For specificity, we again keep the $^{12}$C($n,p$) reaction in mind. The reaction details --- except for its basic kinematics --- are assumed unknown. The reaction itself or its basic details may not even be (properly) implemented in the used simulation package. Therefore, the simulations need to start by generating the exit products (protons), based on the energy and the spatial distribution of the primary reaction-inducing particles (neutrons).

For each separate excited state $\ex$ in the daughter nucleus ($^{12}$B; see table~\ref{tab1}) the \textit{neutron energy} $E$ is sampled from some preselected energy distribution $\hat{\varphi}_{\EE}(\ex,E)$, where we use the hat-notation~$\hat{\cdot}$ to indicate the simulated (as opposed to the later determined, experimental) quantities. These distributions are best selected as uniform or isolethargic, for the simplicity of later analysis. The produced proton direction in the \textit{center-of-mass frame} is then sampled from a preselected angular distribution $\hat{A}(\ex,E,\x)$, which is best selected as isotropic. The proton energy in the center-of-mass frame is calculated based on the \mbox{$Q$-value} for a particular excited state. The proton energy and direction are then boosted into the laboratory frame by the proper (in our case relativistic) transformations. As for the initial proton position, its radial distribution relative to the direction of the neutron beam must be sampled according to the known neutron beam profile; alternatively, the data need to be properly reweighed according to the same beam profile during the later construction of the $\Emx$ matrix. The sampling (or the later data reweighing) of the initial proton position \textit{along} the neutron beam direction depends on the properties of the simulated sample and may vary between extremely simple and rather involved. In case of the thin sample --- implying the combination of the geometric thickness and the total cross section such that \mbox{$\eta\Sigma_\tot(E)\ll1$}, as discussed in a context of eq.~(\ref{w_approx}) --- the longitudinal proton distribution may be sampled uniformly, as the neutron beam attenuation along the sample is negligible. This was the case with our setup. Otherwise, if the beam losses are known to be considerable, then the relative proton production probability along the sample must be properly accounted for. In case of the homogeneous and geometrically regular sample this correction amounts to the factor \mbox{$1-e^{-\eta\Sigma_\tot(E)\times d/D}$} with $d$ as the proton production depth and $D$ as the sample thickness \textit{along the neutron beam direction}; however, this procedure still does not take into account the multiple scattering effects. For more complex samples the correction involves its full spatial characterization.


Each coincidental proton detection by any pair of \mbox{$\Delta \mathrm{E}$-$\mathrm{E}$} strips needs to be recorded by outputting the relevant physical parameters of that particular event. The necessary data consist of: (1)~the primary neutron energy $E$; (2)~the proton emission angle $\x$ from the \textit{center-of-mass frame}; (3)~the unique designation of the activated $\Delta \mathrm{E}$-$\mathrm{E}$ pair of strips; (4)~the energy deposited in those strips. In addition, the excited state $\ex$, the sampled neutron energy distribution $\hat{\varphi}_{\EE}(\ex,E)$ and the proton angular distribution $\hat{A}(\ex,E,\x)$, together with the total number $\hat{\N}_{\EE}(\ex)$ of generated protons within the sampled neutron energy interval $\EE$ also have to be documented for a complete and meaningful utilization of the simulated data. By the virtue of eq.~(\ref{epsilon}), the elements of the design matrix $\Emx$ from eq.~(\ref{Emx_xl}) may be treated as the integrals over the detected counts, so that by identifying the amount $\D^2 \hat{\N}(\ex,E,\x)$ of generated protons as:
\begin{linenomath}\begin{equation}
\D^2 \hat{\N}(\ex,E,\x)= \hat{\N}_{\EE}(\ex) \, \hat{\varphi}_{\EE}(\ex,E)\, \hat{A}(\ex,E,\x)\, \D E \D \x
\label{N_sim}
\end{equation}\end{linenomath}
we may write:
\begin{linenomath}\begin{align}
\big[\Emx\big]_{\alpha\beta}=\int_{E\in\EE}\int_{\x\in[-1,1]} \frac{w(E) P_l(\x)}{\hat{\varphi}_{\EE}(\ex,E)\,\hat{A}(\ex,E,\x)}\frac{\D^2 \hat{N}_\alpha(\ex,E,\x)}{\hat{\N}_{\EE}(\ex)}.
\label{E_int2}
\end{align}\end{linenomath}
This formalism allows us to construct the sought integrals directly on a count-by-count basis, without ever having to build the full coincidental detection probability distributions $\eps_{\alpha}(\ex,E,\x)$, such as those shown in figure~\ref{fig3}. This is achieved simply by taking a weighted sum of all detected counts:
\begin{linenomath}\begin{align}
\big[\Emx\big]_{\alpha\beta}\simeq \frac{1}{\hat{\N}_{\EE}(\ex)} \sum_{q=1}^{\hat{N}_{\alpha\ex}^{(\EE)}} \frac{w(E_q)P_l(\x_q)}{\hat{\varphi}_{\EE}(\ex,E_q)\,\hat{A}(\ex,E_q,\x_q)}.
\label{E_sum2}
\end{align}\end{linenomath}
Here~$\simeq$ symbolically denotes the representation of the integrals from eq.~(\ref{E_int2}), with the index $q$ enumerating all the appropriately detected counts: $\hat{N}_{\alpha\ex}^{(\EE)}$ of them caused by the protons leaving the daughter nucleus in the excited state $\ex$ and being coincidentally detected by the $\alpha$-th pair of strips. In exactly the same manner, the design vector elements from eq.~(\ref{des_vec}) may be expressed as:
\begin{linenomath}\begin{equation}
\epsilon_\alpha^{\,(\EE)} = \sum_{\ex=0}^{\Xmax_\EE}\int_{E\in\EE}\int_{\x\in[-1,1]} \frac{w(E)\,\rho(\ex,E)\,A(\ex,E,\x)}{\hat{\varphi}_{\EE}(\ex,E)\,\hat{A}(\ex,E,\x)} \frac{\D^2 \hat{N}_\alpha(\ex,E,\x)}{\hat{\N}_{\EE}(\ex)}
\end{equation}\end{linenomath}
and thus constructed on a count-to-count basis:
\begin{linenomath}\begin{equation}
\epsilon_\alpha^{\,(\EE)} \simeq \sum_{\ex=0}^{\Xmax_\EE} \frac{1}{\hat{\N}_{\EE}(\ex)}\sum_{q=1}^{\hat{N}_{\alpha\ex}^{(\EE)}} \frac{w(E_q)\,\rho(\ex,E_q)\,A(\ex,E_q,\x_q)}{\hat{\varphi}_{\EE}(\ex,E_q)\,\hat{A}(\ex,E_q,\x_q)},
\label{E_sum3}
\end{equation}\end{linenomath}
where the branching ratios $\rho(\ex,E)$ and angular distributions $A(\ex,E,\x)$ are now taken to be known from an independent source of information.

We remind that the energy deposition cuts used in the analysis of the experimental data are to be implemented precisely at this point, in the construction of the matrix $\Emx$ or the vector $\vec{\epsilon}^{\,(\EE)}$, thus directly affecting the numbers $\hat{N}_{\alpha\ex}^{(\EE)}$ of the acceptable counts. It is also worth noting that the weighting function $w(E)$ is determined by the actual experimental conditions, as opposed to the arbitrary sampling distributions $\hat{\varphi}_{\EE}(\ex,E_q)$ and $\hat{A}(\ex,E_q,\x_q)$. In that, it is evident that both the simulations and the computational procedures from Eqs.~(\ref{E_sum2}) and (\ref{E_sum3}) are immensely simplified when the uniform neutron energy distributions \mbox{$\hat{\varphi}_{\EE}(\ex,E)=1/|\EE|$} and the isotropic proton angular distributions \mbox{$\hat{A}(\ex,E,\x)=1/2$} are used.

\section{Method extension}
\label{extension}

\begin{figure}[b!]
\centering
\includegraphics[width=\width\linewidth]{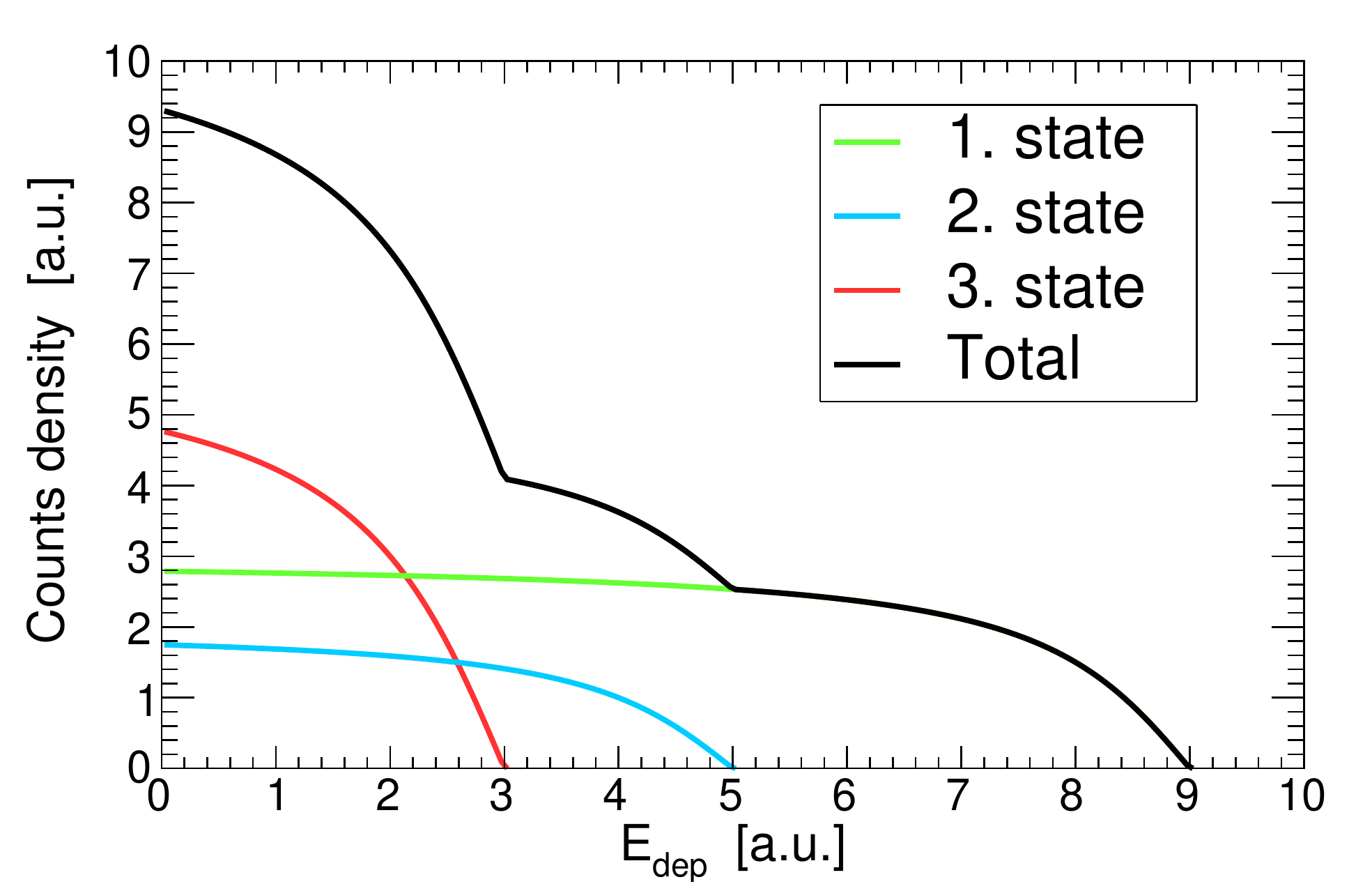}
\caption{Illustrative example: the reaction products leaving the daughter nucleus in any of its excited states may leave a clear signature in the deposited energy spectrum if the energy separation of the excited states is sufficient.}
\label{figx}
\end{figure}

We shortly comment on the possibility of the further method generalization that may be applicable under specific conditions, namely the high statistics and at least a partial separation of the excited states in the deposited energy spectra. For the silicon telescope consisting of $\Delta \mathrm{E}$ and $\mathrm{E}$-layers, the entire two-dimensional \mbox{$\Delta \mathrm{E}$-$\mathrm{E}$} spectra may be considered in the most general case, as we do here. For simplicity, figure~\ref{figx} illustrates the basic idea on the schematic example of the one-dimensional, e.g. \mbox{$(\mathrm{E}+\Delta \mathrm{E})$-spectrum}. Evidently, if the excited states are sufficiently far apart in energy (as determined by the detector resolution), the spectra shapes may serve as an additional source of information to be exploited. In this case one defines the differential coincidental detection probability:
\begin{linenomath}\begin{equation}
\xi_{ij}(\ex,E,\x,\mathrm{E}^{(1)},\mathrm{E}^{(2)})\equiv \frac{\D^4 N_{ij}(\ex,E,\x,\mathrm{E}^{(1)},\mathrm{E}^{(2)})}{\D^2 \N(\ex,E,\x) \D\mathrm{E}^{(1)} \D\mathrm{E}^{(2)}},
\end{equation}\end{linenomath}
starting from the number of counts \mbox{$\D^4 N_{ij}(\ex,E,\x,\mathrm{E}^{(1)},\mathrm{E}^{(2)})$} characterized by the energy $\mathrm{E}^{(1)}$ deposited in the $i$-th \mbox{$\Delta \mathrm{E}$-strip} and the energy $\mathrm{E}^{(2)}$ deposited in the $j$-th \mbox{$\mathrm{E}$-strip}. The master equation for the total number of counts $N_{ij\, \text{\i}\,\text{\j}}^{(\EE)}$ detected within the $\text{\i}$-th \mbox{$\mathrm{E}^{(1)}$-interval} of width $\boldsymbol{\mathsf{E}}_{\text{\i}}^{(1)}$ and the $\text{\j}$-th \mbox{$\mathrm{E}^{(2)}$-interval} of width $\boldsymbol{\mathsf{E}}_{\text{\j}}^{(2)}$ is easily rewritten as:
\begin{linenomath}\begin{align}
\begin{split}
N_{ij\, \text{\i}\,\text{\j}}^{(\EE)}=&\sum_{\ex=0}^{\Xmax_\EE}\int_\EE \D E \int_{-1}^1 \D\x \int_{\boldsymbol{\mathsf{E}}_{\text{\i}}^{(1)}} \D\mathrm{E}^{(1)} \int_{\boldsymbol{\mathsf{E}}_{\text{\j}}^{(2)}} \D\mathrm{E}^{(2)} \,\times\\
&\xi_{ij}(\ex,E,\x,\mathrm{E}^{(1)},\mathrm{E}^{(2)})\,w(E)\,\sigma(E)\,\rho(\ex,E)\,A(\ex,E,\x)
\end{split}
\label{master_ext}
\end{align}\end{linenomath}
where the binning of the deposited-energy distributions, i.e. the set of bin widths $\boldsymbol{\mathsf{E}}_{\text{\i}}^{(1)}$ and $\boldsymbol{\mathsf{E}}_{\text{\j}}^{(2)}$ is entirely arbitrary and may, in the most general case, depend on the particular $(i,j)$-pair of strips and the neutron energy interval $\EE$. In place of the earlier bijective mapping from eq.~(\ref{alpha}), an entire set of indices \mbox{$i,j,\text{\i},\text{\j}$} is now to be mapped onto the unique index $\alpha$:
\begin{linenomath}\begin{equation}
(i,j,\text{\i},\text{\j}) \mapsto\alpha,
\end{equation}\end{linenomath}
allowing the extended design matrix $\Emx$:
\begin{linenomath}\begin{align}
\begin{split}
\big[\Emx\big]_{\alpha\beta}\equiv \int_\EE \,\D E \int_ {-1}^1 \,\D\x \int_{\boldsymbol{\mathsf{E}}_{\text{\i}}^{(1)}} \D\mathrm{E}^{(1)} \int_{\boldsymbol{\mathsf{E}}_{\text{\j}}^{(2)}} \D\mathrm{E}^{(2)}  \times \xi_{ij}(\ex,E,\x,\mathrm{E}^{(1)},\mathrm{E}^{(2)})\,w(E)\,P_l(\x)
\end{split}
\end{align}\end{linenomath}
to be used in bringing eq.~(\ref{master_ext}) to the matrix form from eq.~(\ref{vectorized}), with $\vec{\prt}^{\,(\EE)}$ staying the same as in eq.~(\ref{P_xl}).

\acknowledgments

This work was supported by the Croatian Science Foundation under Project No. 8570.


\end{document}